\documentclass[pre,twocolumn,showpacs,amsmath,amssymb]{revtex4}

\usepackage{graphicx}
\usepackage{bm}

\begin{document}

\title{Heterogeneous animal group models and their group-level
alignment dynamics;\\ an equation-free approach}

\author{Sung Joon Moon$^1$}
\author{B. Nabet$^2$}
\author{Naomi E. Leonard$^2$}
\author{Simon A. Levin$^3$}
\author{I. G. Kevrekidis$^1$}
\email[]{yannis@arnold.princeton.edu}

\affiliation{$^1$Department of Chemical Engineering \&
Program in Applied and Computational Mathematics (PACM),\\
$^2$Department of Mechanical and Aerospace Engineering,\\
$^3$Department of Ecology and Evolutionary Biology,\\
Princeton University, Princeton, NJ 08544, USA}

\begin{abstract}
We study coarse-grained (group-level) alignment
dynamics of individual-based animal group models for
{\it heterogeneous} populations consisting of informed
(on preferred directions) and uninformed individuals.
The orientation of each individual is characterized by
an angle, whose dynamics are nonlinearly coupled with
those of all the other individuals,
with an explicit dependence on the difference between
the individual's orientation and the instantaneous
average direction.
Choosing convenient coarse-grained variables (suggested by
uncertainty quantification methods) that account for rapidly
developing correlations during initial transients, we perform
efficient computations of coarse-grained steady states and
their bifurcation analysis.
We circumvent the derivation of coarse-grained governing
equations, following an equation-free computational approach.
\end{abstract}


\maketitle

\section{Introduction}
 \label{intro}

Coordinated motions and pattern formation have been studied for
a wide range of biological organisms, from bacteria and amoebae
to fish, from birds and wildebeest to humans (Ben-Jacob et al.,
2000; Deneubourg et al., 2001; Parrish et al., 2002; Partridge,
1982; Wilson, 1975).
Animal groups often behave as if they have a single mind,
displaying remarkable self-organized behavior.
At one extreme, the individuals seem to need little information
transfer (e.g. fish schools), while at the other end the
information exchange occurs in highly integrated ways through
long-term associations among the individuals (e.g. honeybee hives
and human communities).
Controlling such an organized behavior in groups of artificial
objects, including autonomous underwater vehicles
(Leonard et al., 2006) and groups of autonomous agents
(Jadbabaie et al., 2003), has received extensive attention
in contemporary control theory.
Challenges in both natural and engineering settings involve
understanding which patterns emerge from the interaction
among individual agents.

Select laboratory experiments have shed some light on the
schooling mechanism (Hunter, 1966; van Olst and Hunter, 1970;
Partridge and Pitcher, 1979, 1980; Pitcher et al., 1976).
It still remains unclear, however, how the individual-level
behavior and group-level (``macroscopic'', or coarse-grained)
patterns are related.
More precise experiments using three-dimensional tracking of
every individual in a population should lead to better
understanding of this linkage.
An ultimate experimental study with precise control of every
relevant detail may not be possible, yet appropriate
mathematical models would provide a venue to establish
behavioral cause, as one can consider different hypothetical
individual-level interaction rules selectively
(see e.g., Flierl et al., 1999).

Several different individual-based models have been proposed,
which reproduce certain types of collective behavior in animal
groups (e.g., see Aoki, 1982; Reynolds, 1987;
Deneubourg and Goss, 1989).
Self-organization emerges also in a wide spectrum of physical
and chemical systems, some of which (e.g., crystals and
ferromagnetic materials) exhibit apparent similarities with
emergent patterns observed in animal groups.
Vicsek et al. (1995) have introduced a discrete-time
model of self-driven particles, or self-propelled particles (SPP),
based on near-neighbor rules that are similar with those
in the ferromagnetic XY model (Kosterlitz and Thouless, 1973).
The authors analyzed statistical properties of the model,
including phase transition and scaling (Vicsek et al., 1995).
A long-range interaction has been incorporated into the SPP
model (Mikhailov and Zanette, 1999), and continuum, ``hydrodynamic''
versions of this model have been introduced (Toner and Tu, 1995,
1998; Topaz et al., 2006).
Recently, Couzin et al. (2002, 2005) have introduced a model
to provide insights into the mechanism of decision making
in biological systems, which reproduces many important
observations made in the field, and provides new insights
into these phenomena.
A review for various models can be found in Parrish et al. (2002)
and Czir{\'o}k and Vicsek (2001).

The models of Couzin et al. (2005), and most other such models,
incorporate various detailed mechanistic steps.
These shed light on the role of leadership and imitation, and
produce a number of surprising results, such as the influence
that a few ``informed'' individuals can have on large collectives.
What is needed now are efforts to simplify those models, and
to show especially what properties of the microscopic simulators
are essential to explain that behavior.
For some models, closure schemes are available (Flierl et al., 1999);
but more generally, though we may suspect that closures exist,
we cannot derive explicit expressions for them.
In such circumstances, we need methods such as those used in
this paper; we perform the coarse-grained dynamical analysis
by circumventing the derivation of governing equations, using
an equation-free computational approach (Theodoropoulos et al.,
2000; Kevrekidis et al., 2003).
A particular goal is to understand how much of the specific
spatial detail is fundamental to the behavior.
But turning to the Kuramoto-type approximation, where the
interaction is assumed to be global,
we deliberately ignore local effects.
To the extent that the models fail to explain observed types of
behavior, we will need to turn next to more detailed models. 

Most of previously proposed models concern populations of
{\it homogeneous} (or indistinguishable) individuals.
Furthermore,
the dynamical analysis in the literature is often limited to
a small subset of the entire parameter space, and a systematic
classification of possible global dynamics remains elusive.
In the current paper, we study the {\it coarse-grained}
alignment dynamics of individual-based animal group models.
The measurement of the mean angular deviation of fish schools
(e.g. clupeids and scombroids; see Atz, 1953; Hunter 1966)
showed that it varies continuously from no alignment to
practically perfect alignment.
We account for this continuous change by heterogeneity
(``quenched noise''; characterized by parameters of random
variables drawn from
a prescribed distribution function) and the coupling strength.
Our approach is flexible in that the heterogeneity can
be introduced in various places in the model, and 
the way we analyze different heterogeneity cases does not
require any significant modification.

The rest of the paper is organized as follows:
Models for homogeneous and heterogeneous animal groups are
described in Secs.~\ref{model1} and \ref{model_extension},
and our approach, equation-free polynomial chaos, is explained
in Secs.~\ref{method1} and \ref{method2}.
Coarse-grained dynamical analysis and its comparison with
fine-scale dynamics, for a system of two informed individuals
and a large number of heterogeneous uninformed individuals,
are presented in Sec.~\ref{results}.
The case of two groups of heterogeneous informed individuals
is presented in Sec.~\ref{results2}.
We conclude with a brief discussion in Sec.~\ref{conclusions}.

\section{Models and methods}
 \label{modelmethod}

\subsection{A ``minimal'' model for identical individuals}
 \label{model1}

We briefly discuss a ``minimal'' model proposed by Nabet
et al. (2006), which we extend in our study.
It concerns the alignment dynamics of a homogeneous population
of {\it indistinguishable} $N$ individuals with two subgroups
of informed individuals (``leaders'') with populations $N_1$
and $N_2$ respectively and $N_3$ uninformed individuals
(``followers''), where $N = N_1 + N_2 + N_3$:
\begin{eqnarray}
{d\psi_1 \over dt} & = & \sin(\Theta_1 - \psi_1) + \frac{K}{N} N_2 \sin(\psi_2 - \psi_1) + \frac{K}{N} N_3 \sin(\psi_3 - \psi_1), \nonumber \\
{d\psi_2 \over dt} & = & \sin(\Theta_2 - \psi_2) + \frac{K}{N} N_1 \sin(\psi_1 - \psi_2) + \frac{K}{N} N_3 \sin(\psi_3 - \psi_2), \nonumber \\
{d\psi_3 \over dt} & = & \frac{K}{N} N_1 \sin(\psi_1 - \psi_3) + \frac{K}{N} N_2 \sin(\psi_2 - \psi_3).
\label{reduced}
\end{eqnarray}
%
Here $\psi_k$ characterizes the average direction of the
individuals in each of the two informed subgroups for $k=1,2$
and the average direction of the uninformed individuals for $k=3$.
$\Theta_k$ is the corresponding informed, preferred direction
($\Theta_1$ can be set to zero without loss of generality)
and $K (\geq 0)$ is the coupling strength.
This minimal model corresponds to the reduced system of the
following system of $N$ individuals (Nabet et al., 2006):
\begin{eqnarray}
{d\theta_j \over dt} & = & \sin(\Theta_1 - \theta_j) + \frac{K}{N} \sum_{l=1}^N \sin(\theta_l - \theta_j),~~1\leq j \leq N_1,   \label{m1} \nonumber \\
{d\theta_j \over dt} & = & \sin(\Theta_2 - \theta_j) + \frac{K}{N} \sum_{l=1}^N \sin(\theta_l - \theta_j),~~N_1+1\leq j \leq N_1 + N_2,   \label{m2} \nonumber \\
{d\theta_j \over dt} & = &  \frac{K}{N} \sum_{l=1}^N \sin(\theta_l - \theta_j),~~~~~~~~~~~~~N_1+N_2+1\leq j \leq N, \label{m3}
\label{nabet_full}
\end{eqnarray} 
%
where the angle $\theta_j$ characterizes the direction in which
the $j$th individual is heading (we will refer to it as ``orientation'').
The average direction $\psi_k$ is defined as the angle of the
average of the phasors (when each individual's dynamical state
is considered as a phasor of unit radius and a phase angle)
of the individuals in the $k$th subgroup;
$\rho_k$ is the magnitude of the average of the phasors.
Formally, this is written as
\begin{eqnarray}
\rho_1 e^{i\psi_1} &\equiv& \frac{1}{N_1} \sum_{j=1}^{N_1} e^{i\theta_j}, \nonumber \\
\rho_2 e^{i\psi_2} &\equiv& \frac{1}{N_2} \sum_{j=N_1 +1}^{N_1+N_2} e^{i\theta_j}, \nonumber \\
\rho_3 e^{i\psi_3} &\equiv& \frac{1}{N_3} \sum_{j=N_1+N_2 +1}^{N} e^{i\theta_j}. \nonumber
\end{eqnarray}
The large population model
in Eq.~(\ref{m3}) has a separation of time scales.
Individuals within each subgroup synchronize quickly, i.e.,
$\rho_k$ quickly converges to 1.
The slow dynamics are described by the reduced system
(Eq.~(\ref{reduced})) where the variables $\psi_k$ characterize
the {\it lumped behavior} of each of the three subgroups.

It is assumed that the alignment (orientational) dynamics are
independent of the translational counterpart (Sepulchre et al., 2005);
hence, the dynamical
state of an individual can be characterized by its orientation.
The functional form for mutual interaction is borrowed from
the well-known Kuramoto model (Kuramoto, 1984), a prototypical
model for coupled nonlinear oscillators.
This simplified global interaction model is consistent with an
observation that the strongest correlations are observed between the
(speed and) direction of the individual and the average (speed
and) direction of the entire school (Partridge, 1982):
In the mean-field form of the Kuramoto model, the interaction
term can be rewritten as
\begin{equation}
{1\over N}\sum_{l=1}^N \sin(\theta_l-\theta_j) = r\sin(\psi-\theta_j),
\end{equation}
where
\begin{equation}
re^{i\psi} \equiv {1\over N}\sum_{l=1}^N e^{i\theta_l}. \nonumber
\end{equation}
In this alternate expression, the dependence on the difference
between the individual direction and the average direction
stands out explicitly. 
In the absence of coupling ($K=0$), each leader eventually
heads for its preferred direction.
Nontrivial dynamical behavior for the minimal model
(Eq.~(\ref{reduced})) are studied in Nabet et al. (2006);
bifurcations are analyzed for the global phase space in
the case $N_1 = N_2$ and $N_3 = 0$.

\subsection{Extension to heterogeneous populations}
 \label{model_extension}

The aforementioned models concern populations of
{\it homogeneous} subgroups, where the individuals in each
subgroup quickly synchronize, nearly perfectly ($\rho_k \sim 1$),
during the initial transients (Nabet et al., 2006).
In the more general case, the mean angular deviation of fish
schools is finite (Atz, 1953; Hunter, 1966), which is not
captured in this ``minimal'' model.
We extend the model to account for the distribution of
directions within schools, assuming it arises from the
{\it heterogeneity} among the group members.
We introduce the heterogeneity in the following two ways:
\\

(I) {\it Two leaders and many heterogeneous followers ---}
First we consider the cases when the population consist of
two leaders (which possibly represent lumped behavior of
groups of homogeneous leaders) and $N~(\gg 1)$ followers:
\begin{widetext}
\begin{eqnarray}
 \label{microEQ1}
{d\psi_1\over dt} &=& \sin(\Theta_1-\psi_1) + {K\over N+2}\left[\sum_{j=1}^{2}\sin(\psi_j-\psi_1)+\sum_{j=1}^{N}\sin(\theta_j-\psi_1)\right], \nonumber \\
{d\psi_2\over dt} &=& \sin(\Theta_2-\psi_2) + {K\over N+2}\left[\sum_{j=1}^{2}\sin(\psi_j-\psi_2)+\sum_{j=1}^{N}\sin(\theta_j-\psi_2)\right], \\
{d\theta_i\over dt} &=& \omega_i + {K\over N+2}\left[\sum_{j=1}^{2}\sin(\psi_j-\theta_i)+\sum_{j=1}^{N}\sin(\theta_j-\theta_i)\right]~~~~~~~~~~~~~~~{\rm for}~1\leq i \leq N, \nonumber
\end{eqnarray}
\end{widetext}
where the heterogeneity is accounted for through the tendency
to deviate from the average direction, characterized
by $\omega_i$, an i.i.d. random variable drawn from a
prescribed distribution function $g(\omega)$
(of standard deviation $\sigma_{\omega}$ with mean value zero).
For notational convenience, we drop a subscript of
a variable to represent a random variable of a proper length
({\it cf.} $\omega_i$ and $\omega$).
As $\Theta_1$ can be set to zero without loss of generality,
$\Theta_2$ and $K$ are control parameters.
In the current study, we consider $g(\omega)$ to be Gaussian,
but our analysis is not limited to this particular choice.
\\

(II) {\it Two groups of heterogeneous leaders ---}
Secondly, we consider two groups of heterogeneous leaders without
any followers, focusing only on the dynamics among leaders.
The heterogeneity is accounted for by introducing randomness
in the angles preferred by the leaders.
The orientations of the leaders in each group are denoted by
$\chi_i$'s and $\phi_i$'s (of sizes $N_1$ and $N_2$) respectively:

\begin{widetext}
\begin{eqnarray}
 \label{for_many}
{d\chi_i\over dt} &=& \sin({\mathcal X}_i-\chi_i) + {K\over N_1 + N_2}\left[\sum_{j=1}^{N_1}\sin(\chi_j-\chi_i) + \sum_{j=1}^{N_2}\sin(\phi_j-\chi_i)\right] {\rm ~~~~~for}~1 \leq i \leq N_1, \nonumber \\
{d\phi_i\over dt} &=& \sin(\Phi_i-\phi_i) + {K\over N_1 + N_2}\left[\sum_{j=1}^{N_1}\sin(\chi_j-\phi_i) + \sum_{j=1}^{N_2}\sin(\phi_j-\phi_i)\right]~ {\rm ~~~~~for}~1 \leq i \leq N_2,
\end{eqnarray}
\end{widetext}
where the preferred angles ${\mathcal X}_i$ and $\Phi_i$ are randomly
drawn from prescribed distributions $g_1({\mathcal X})$ and $g_2(\Phi)$
(i.e., i.i.d. random variables of standard deviations
$\sigma_{{\mathcal X}}$ and $\sigma_{\Phi}$), respectively.
We set $<{\mathcal X}> = 0$, and will vary $K~(\geq 0)$ and $<\Phi>$
$(\in [0,\pi])$ as control parameters (and investigate some cases
of different values of $\sigma_{{\mathcal X}}$ and $\sigma_{\Phi}$ in
Sec.~\ref{results2b}).

\begin{figure}[t]
\begin{center}
\includegraphics[width=.32\columnwidth]{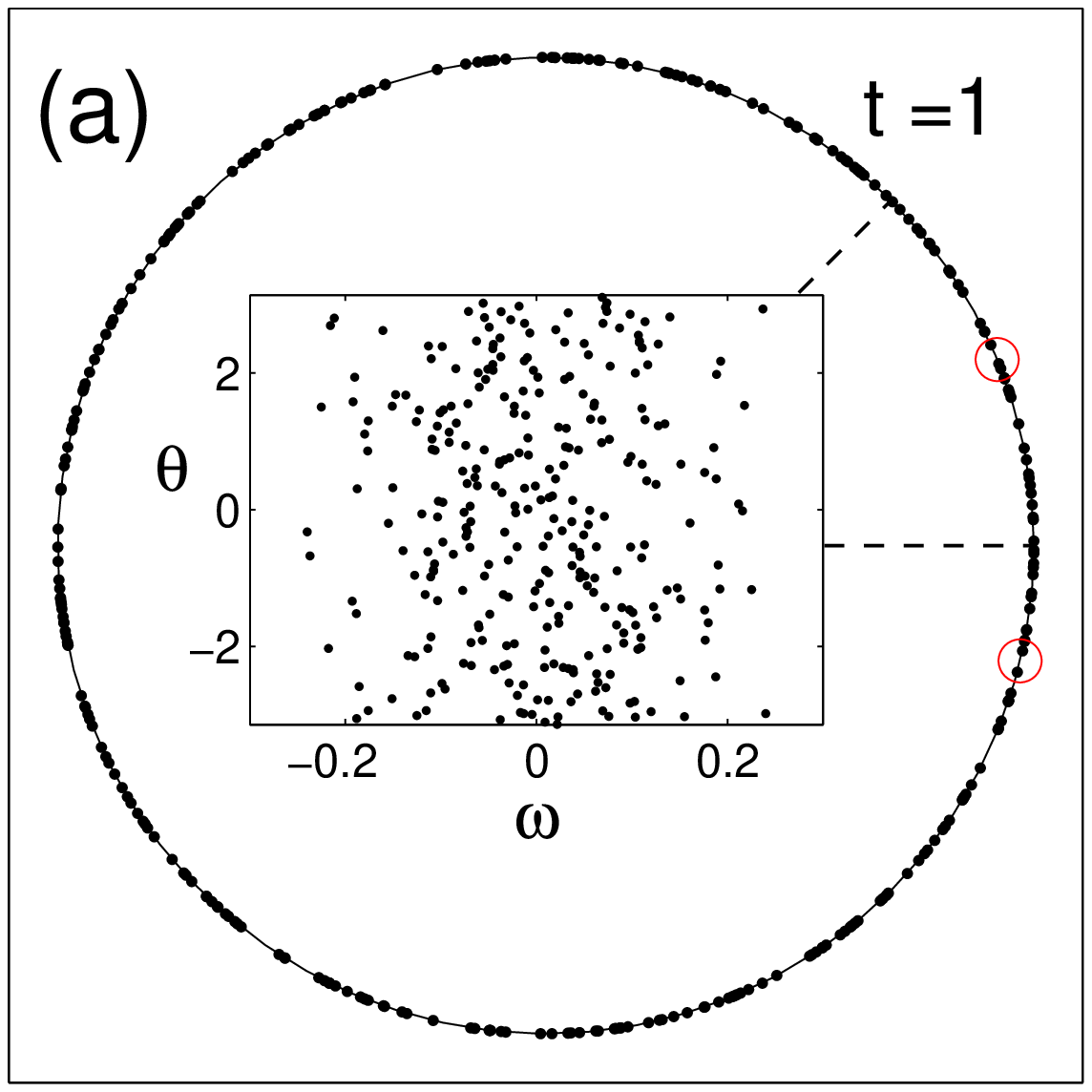}
\includegraphics[width=.32\columnwidth]{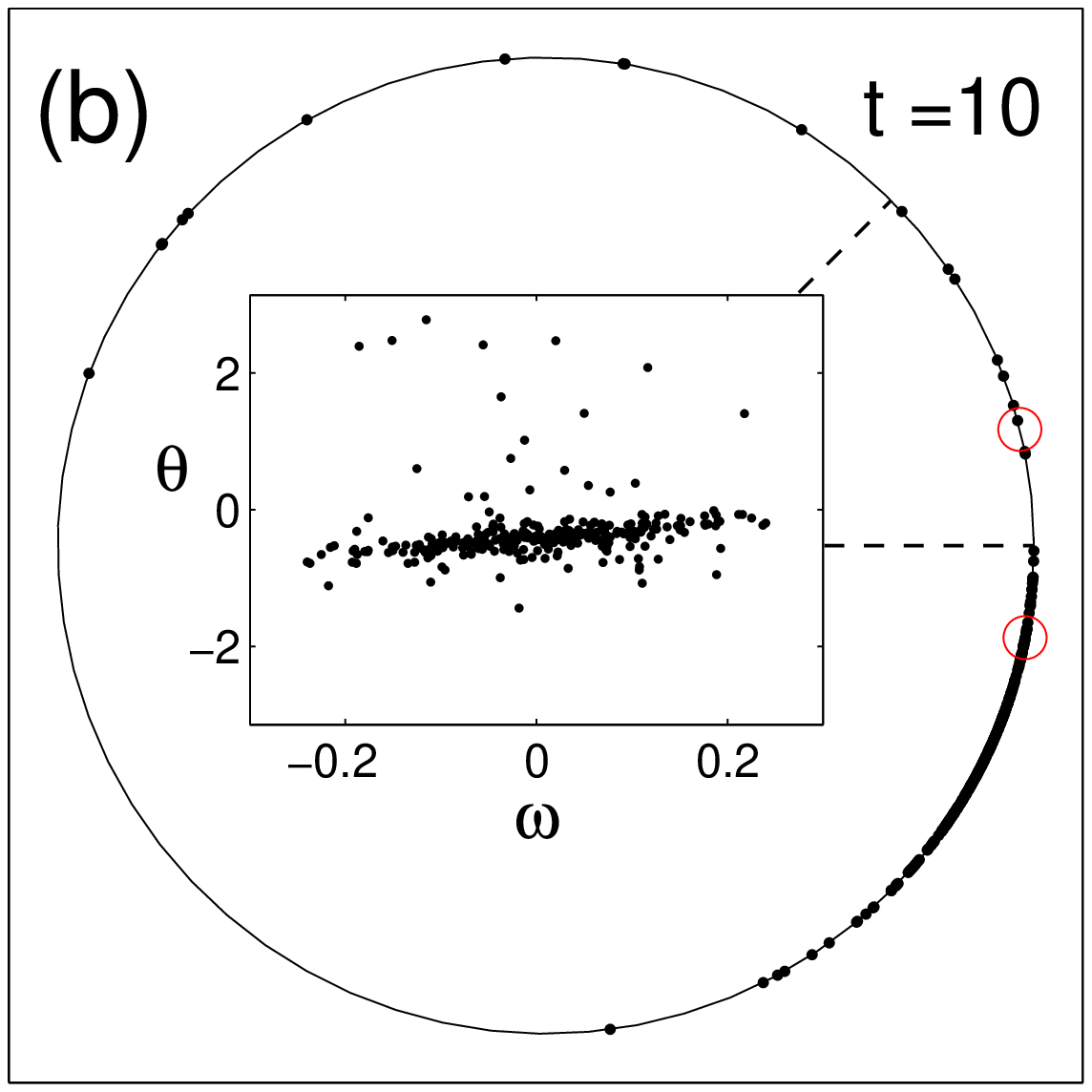}
\includegraphics[width=.32\columnwidth]{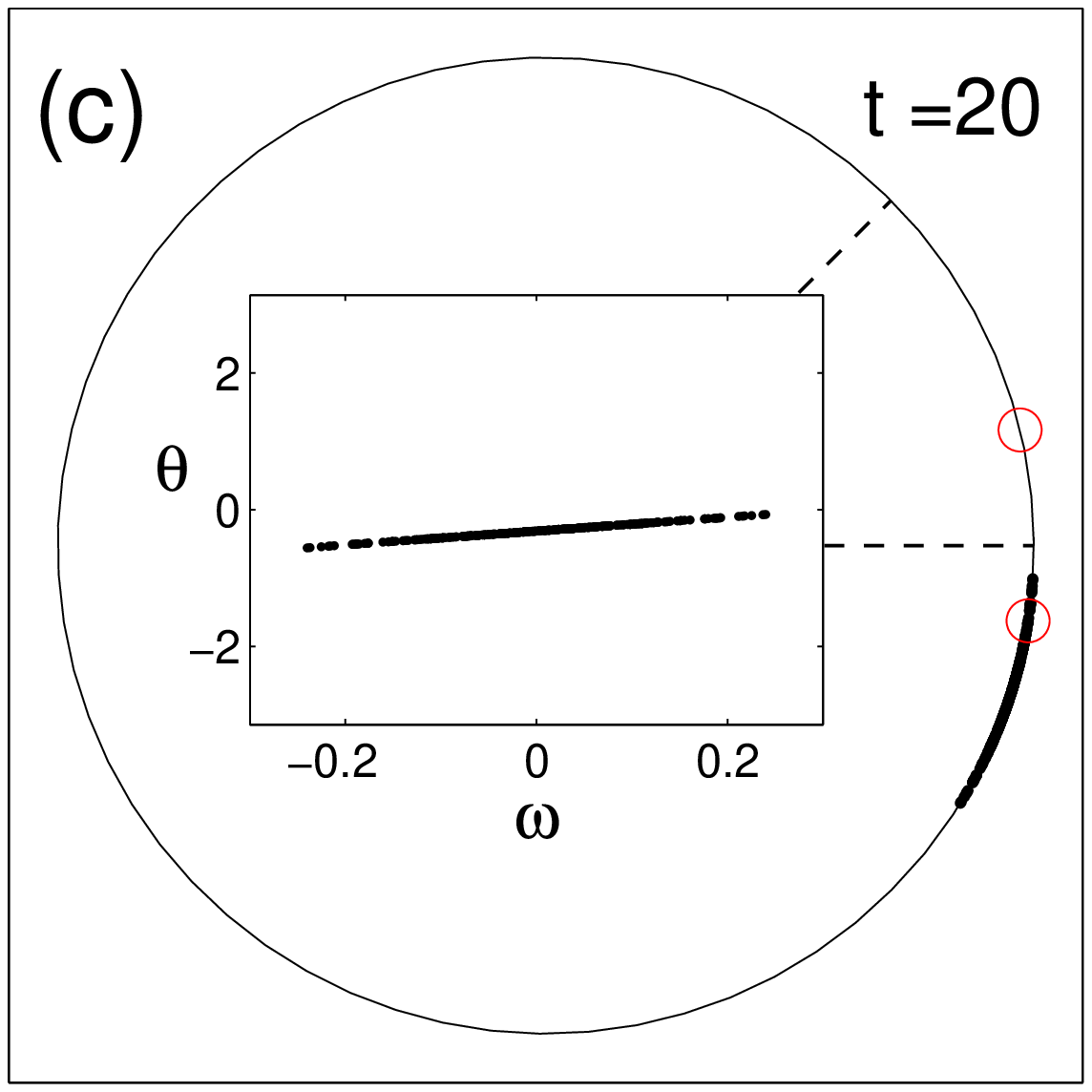}

\includegraphics[width=.32\columnwidth]{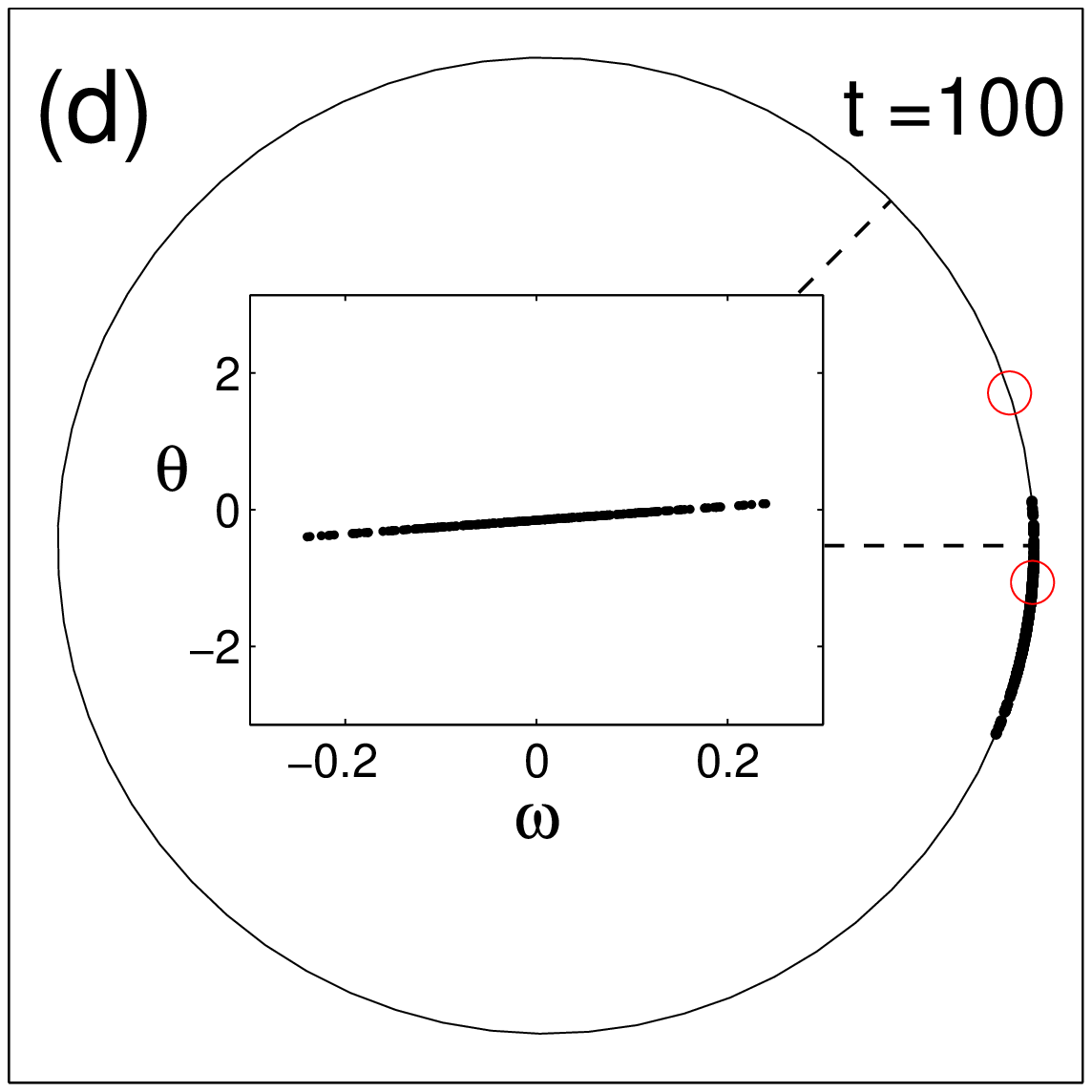}
\includegraphics[width=.32\columnwidth]{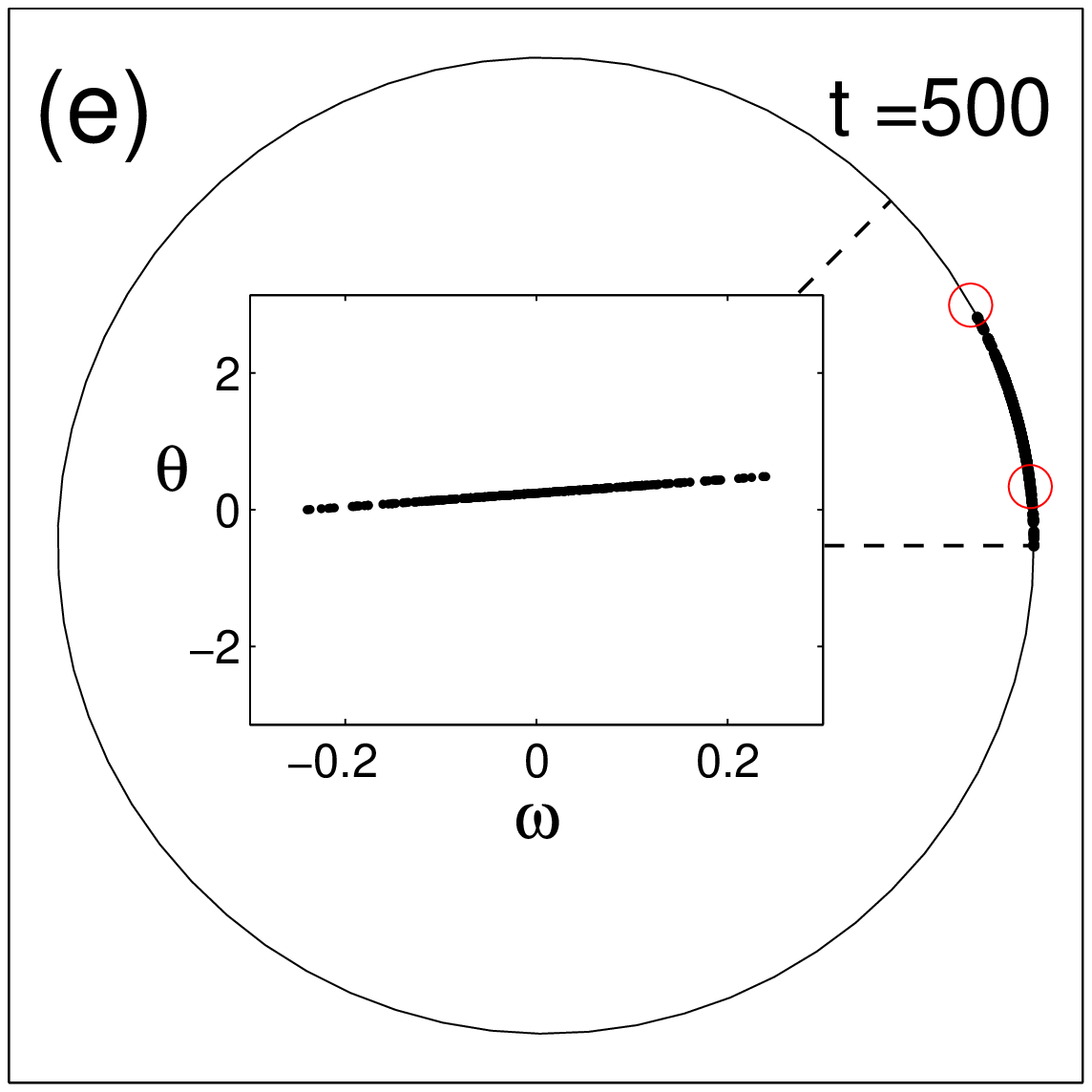}
\includegraphics[width=.32\columnwidth]{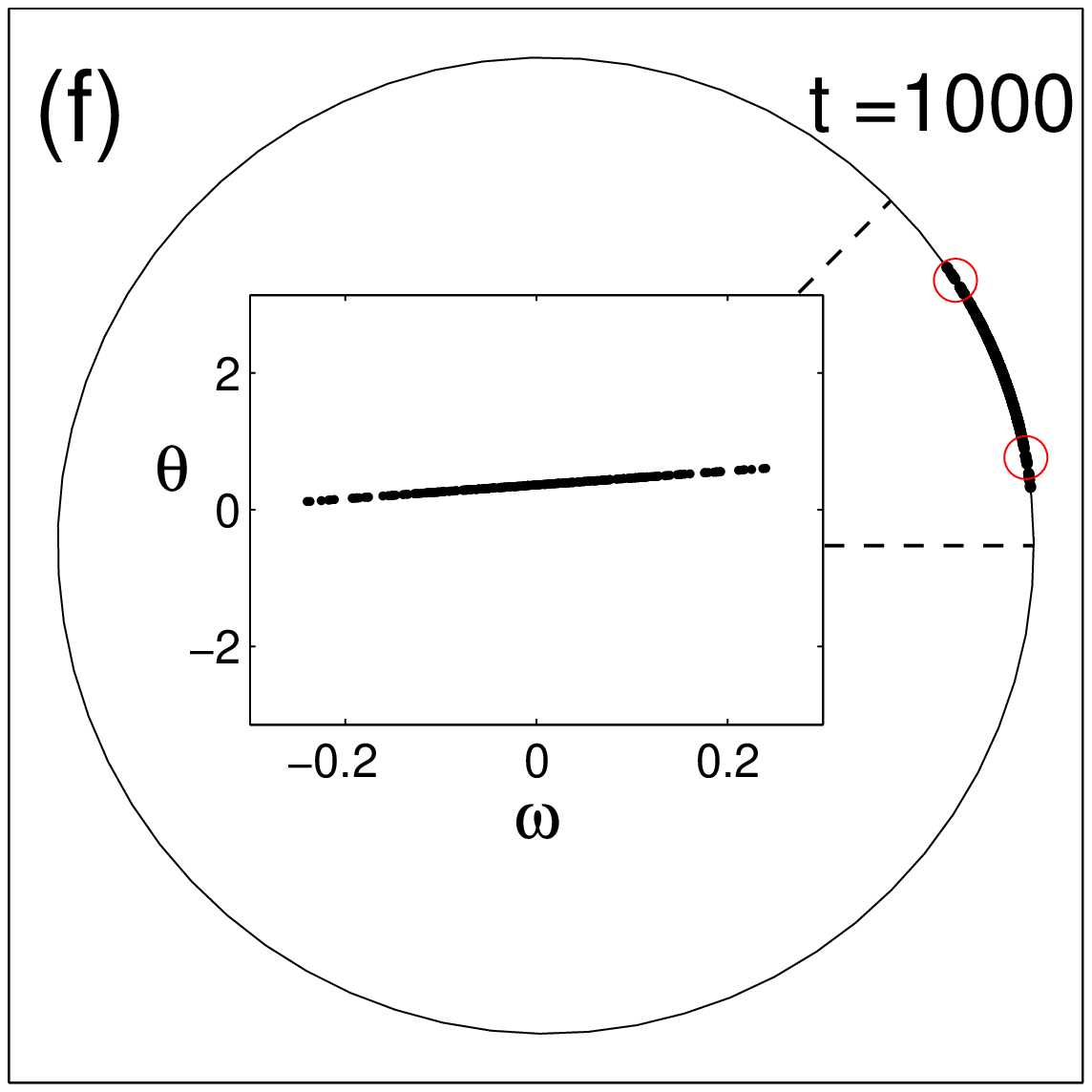}
\caption{
\label{correlation}
Direct integration of a system of two leaders (open circles;
dashed lines indicate preferred angles) and 300 followers
(dots), initialized from uniformly distributed orientations
with randomly assigned heterogeneity variable (i.e.,
no initial correlations between $\theta$ and $\omega$),
is shown for an initial transient [(a) to (c)],
and for much longer time scales [(d) to (f)].
Insets illustrate time evolution of the followers' orientations
on the $\theta-\omega$ plane, where strong correlations develop
during a short time $t \sim 10$.
After that, the leaders and followers, the latter effectively
as a ``unit'', slowly drift to the stable steady state.
It takes of the order of $t \sim 10^3$ for the system to
asymptotically converges to this final state.
($K = 1.0;~\Theta_2 = \pi/4$).
}
\end{center}
\end{figure}

\subsection{Wiener polynomial chaos}
 \label{method1}

The Kuramoto model, a paradigm for all-to-all, phase-coupled
oscillator models, has been extensively studied and used to
shed light on many synchronization
phenomena (Kuramoto, 1984; Acebr{\'o}n et al., 2005, and
references therein).
This model has the property that, in the full synchronization
regime (of large enough $K$ values), phase angles become quickly
correlated with (or ``sorted" according to) the natural frequencies
during the initial short transients (Moon et al., 2006).
Similar correlations (between the angles and heterogeneity random
variables) are expected to arise in the current
model (which is indeed the case, as will be shown later in
Fig.~\ref{correlation}), since the coupling term is qualitatively
similar.
As in Moon et al. (2006), we choose expansion coefficients
in Wiener polynomial chaos as coarse-grained ``observables'',
to explore low-dimensional, coarse-grained dynamics.

Wiener(-Hermite) polynomial chaos was introduced by Wiener (1938),
who represented a random process in terms of functional expansions
of Wiener process (historically, this method has been termed as
polynomial ``chaos'', because of its initial usage on homogeneous
chaos, such as turbulence and Brownian motion, rather than the
nature of the method).
Ghanem and Spanos (1991) later extended this idea to treat
random processes as functional expansions of random variables,
or elements in the Hilbert space of random functions, in which
a spectral representation in terms of polynomial chaos is
identified.
The projections (or coefficients) on the polynomial base then
can be determined through a Galerkin approach.
This method was subsequently applied in uncertainty
quantification of various problems (e.g., see
Ghanem, 1999; Jardak et al., 2002), and has been extended to
general situations using the Askey scheme (Xiu et al., 2002;
now known as generalized polynomial chaos).

In this method, dependent random variables ($\theta$ of the
followers for the case (I), and $\chi$ and $\phi$ for the case
(II)) are expanded in polynomials of independent random variables
($\omega$, or ${\mathcal X}$ and $\Phi$) using appropriately chosen basis
functions.
Details for the two cases are as follows:
\\

(I) {\it Two leaders and many heterogeneous followers ---}
For convenience, we introduce the unit Gaussian random
variable $\xi \equiv \omega/\sigma_{\omega}$.
Using this newly defined variable, we expand $\theta(\omega,t)$
(i.e. $\theta(\xi,t)$) in Hermite polynomials of $\xi$
[$H_0(\xi)$ = 1, $H_1(\xi) = \xi$, $H_2(\xi) = \xi^2- 1$,
$H_3(\xi) = \xi^3 - 3\xi, \cdots$]:
\begin{equation}
 \label{expansion}
\theta(\xi,t) = \sum_{n=0}^p \alpha_n(t) H_n\left(\xi\right),
\end{equation}
where $p$ is the highest order retained in the truncated series,
$H_n$ is the $n$th Hermite polynomial, and the $\alpha_n$'s are
the expansion coefficients which will be referred to simply
as ``chaos coefficients'' in this paper.
Wiener polynomial chaos, utilizing Hermite polynomials as basis
functions, is the appropriate choice for Gaussian random
variables that we consider in the present study.
The probability density function of the Gaussian random
variables appears as the weighting function of Hermite
polynomials, and the Hermite polynomial expansion is suggested
to converge exponentially for Gaussian processes (Lucor et al., 2001).
For other random variables, use of different basis functions
(for instance, Legendre polynomials for uniform random variables)
has been suggested for fast convergence, which is the basis of
the development of the generalized polynomial chaos (Xiu et al., 2002).

We choose the first few nonvanishing chaos coefficients $\alpha_n$'s,
as well as the orientations of the leaders ($\psi_1$ and $\psi_2$),
to be the coarse-grained ``observables''.
Due to symmetry, all the even order $\alpha_n$'s vanish,
except for the zeroth order $\alpha_0$ that corresponds to
the average direction of the followers.
Geometrically, $\alpha_1$ and $\alpha_3$ respectively represent
a measure for the linear order spread of the angles (the
``slope'' between $\theta$ and $\omega$) and the cubic order
measure.
In the continuum limit ($N \rightarrow \infty$), the chaos
coefficients can be exactly determined using the orthogonality
relations for Hermite polynomials.
However, in the {\it finite} cases of single realization we
consider, $N \sim O(10^2)$, those relations hold only approximately,
and the coefficients are evaluated using least squares fitting,
following Moon et al. (2006).
\\

(II) {\it Two groups of heterogeneous leaders ---}
In the second case, we expand $\chi$ and $\phi$ in terms of
${\mathcal X}$ and $\Phi$, respectively:
\begin{eqnarray}
 \label{expansion2}
\chi &=& \sum_{n=0}^{p} \alpha_n H_n(\zeta), \nonumber \\
\phi &=& \sum_{n=0}^{p} \beta_n H_n(\eta),
\end{eqnarray}
where the chaos coefficients \mbox{\boldmath $\alpha$}
and \mbox{\boldmath $\beta$} are the coarse ``observables''
of our choice, $H_n$'s are Hermite polynomials (for Gaussian
$g_1$ and $g_2$), and $\zeta \equiv {\mathcal X}/\sigma_{\chi}$ and
$\eta \equiv \Phi/\sigma_{\phi}$ are unit Gaussian random
variables.
\\

\subsection{``Equation-free'' computational approach}
 \label{method2}

A prerequisite to {\it coarse-grained} dynamical analysis (which
is the main goal of the current study) is, in a traditional sense,
an explicit derivation of coarse-grained governing equations.
In principle, such equations for chaos coefficients, in the
continuum limit ($N \rightarrow \infty$), might be obtained
through a stochastic Galerkin method (Ghanem and Spanos, 1991).

In the present study, we do not even attempt to derive such
equations.
We circumvent their derivation by using an equation-free
multiscale computational
approach (Theodoropoulos et al., 2000; Kevrekidis et al.,
2003, 2004).
This approach enables us to explore the coarse-grained
dynamics without the assumption of the continuum limit.
The premise of this approach is that coarse-grained governing
equations conceptually {\it exist}, but are not explicitly
available in closed form.
This approach is based on the recognition that short bursts of
appropriately initialized microscopic (fine-scale) simulations
during a time horizon $\Delta T$ and the projection of the results
onto coarse-grained variables, say \mbox{\boldmath $x$}, result in
time-steppers (mappings) for those variables ${\mathbf \Phi}_{\Delta T}$
(which is effectively the same as the discretization of unavailable
equations):
\begin{equation}
 \label{coarse_stepper}
\mbox{\boldmath $x$}_{n+1} = {\mathbf \Phi}_{\Delta T}(\mbox{\boldmath $x$}_n).
\end{equation}
One then processes the results of the short simulations to estimate
various coarse-grained quantities (such as time derivatives,
action of Jacobians, residuals) to perform relevant coarse-grained
level numerical computations, as if those quantities were obtained
from coarse-grained governing equations.
For instance, one can {\em integrate} unavailable governing equations
in time (coarse projective integration; see below), or compute the
steady states of the above coarse time-stepper, by utilizing
fixed point algorithms (such as Newton-Raphson or Newton-GMRES).

Equation-free computations consist of the following steps:
\begin{itemize}
\item[1.] Identify coarse-grained variables (``coarse
observables'') that sufficiently describe both the micro- and
macroscopic dynamics; in our study, they are $\alpha_n$'s
(and $\beta_n$'s).
For convenience, we denote the microscopic (macroscopic)
descriptions by $\mathbf{\theta}$ ($\mathbf{\alpha}$).
\item[2.] Choose an appropriate {\it lifting} operator $\mu_L$,
which maps $\mathbf{\alpha}$ to one (or more) consistent
description(s) $\mathbf{\theta}$ (for the purposes of variance
reduction and ensemble-averaging).
In the current study, this can be achieved by using the relations
in Eqs.~(\ref{expansion}) and (\ref{expansion2}); once random
variables are drawn, these relations are used to obtain
corresponding $\mathbf{\theta}$.
\item[3.] Starting from lifted initial condition(s)
$\mathbf{\theta}(t_0) = \mu_L(\mathbf{\alpha}(t_0))$,
run the microscopic simulator to obtain $\mathbf{\theta}(t_0+\Delta T)$
at a later time ($\Delta T \geq 0$).
\item[4.] Use an appropriate {\it restriction} operator $\mathcal{M}_R$
(least squares fitting, in the current study)
which maps the microscopic state(s) to the macroscopic description
$\mathbf{\alpha}(t_0+\Delta T) = \mathcal{M}_R(\mathbf{\theta}(t_0+\Delta T))$,
which effectively results in time series of coarse observables, or their
coarse time-stepper ${\mathbf \Phi}_{\Delta T}$;
$\mathbf{\alpha}(t_0+\Delta T) \equiv {\mathbf \Phi}_{\Delta T}(\mathbf{\alpha}(t_0))$.
\item[5.] Apply desired numerical techniques using the coarse-grained
variables obtained from the step 4. and repeat some of the above
steps as needed.
\end{itemize}
An extensive discussion can be found in Kevrekidis et al. (2003, 2004).

\section{Results for Case I}
 \label{results}

Direct integration of the ``fine-scale'' model of
Eq.~(\ref{microEQ1}) in the strong coupling regime ($K = 1.0,
~\sigma_{\omega} = 0.1$), started from randomly assigned orientations
and the heterogeneity variable (the latter is a Gaussian random
variable), illustrates that a strong correlation between $\theta$
and $\omega$ develops during a short, initial transient time;
the orientations of the followers quickly become a monotonically
increasing function of their heterogeneity variable (Fig.~\ref{correlation}),
after which they slowly drift as a ``unit'' until they settle
down in the final steady state.
During the latter slow drift, the system can be described as two
leaders and a {\it single} ``clump'' of followers, whose
coarse-grained states can be successfully described by a small
number of chaos coefficients.
A similar time scale separation exists in the model of homogeneous
populations.
In this case, followers quickly collapse asymptotically to the
{\it same} direction (Nabet et al., 2006).

\begin{figure}[t]
\begin{center}
\includegraphics[width=.62\columnwidth]{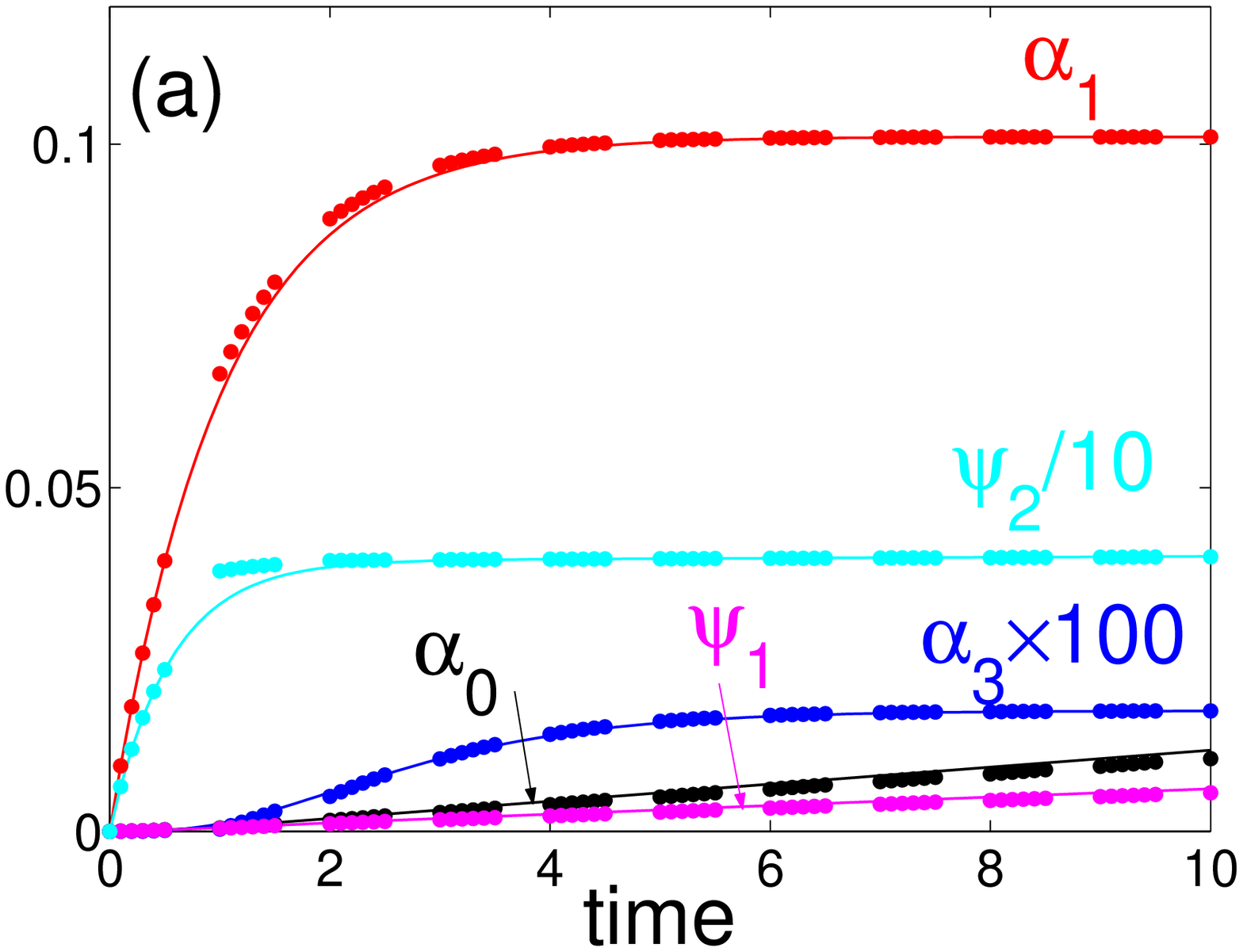}
\includegraphics[width=.6\columnwidth]{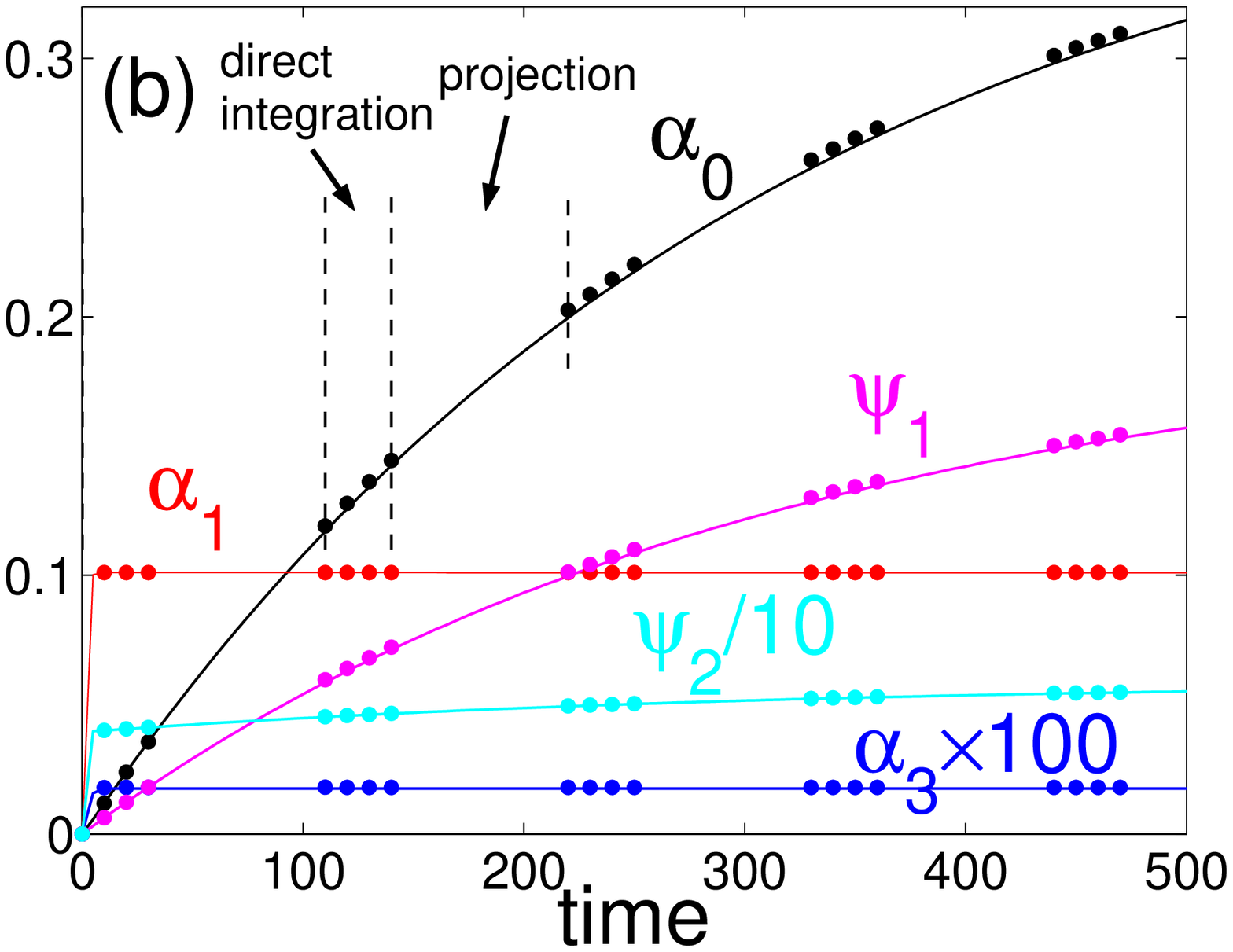}
\caption{
\label{PI}
(Color online)
Accelerated computation of stable steady states via coarse
projective integration using five coarse-grained variables,
shown here for two different time scales
($K = 1.0;~\Theta_2 = \pi/4$).
Initially all the values are assigned to be 0.
Both $\alpha_1$ and $\alpha_3$ reach their steady state values
relatively quickly (see (a)), while the others are slowly
varying (see (b); they are still varying at $t = 500$).
Dots represent the time intervals during which short direct
integration is performed (and restricted), in the course of
the projective integration using forward Euler method.
Solid lines represent the trajectories of direct full integration
during the entire time.
Higher efficiency can be achieved by optimally choosing the
time horizon for the direct integration, the projection
stepsize, and projection method.
}
\end{center}
\end{figure}

\subsection{Computations of steady states}

We begin by accelerating the approach to a stable steady state
using an equation-free algorithm, the coarse projective
integration method (Gear and Kevrekidis, 2003).
In contrast to a conventional, direct integration of the full
fine-scale model during the {\it entire} time (until sufficient
convergence to stable, final states), this method exploits
smoothness in the coarse variables (estimated through a direct
integration during a {\it short} time), in order to extrapolate
and take a large projective time-step (compared to the original
integration time-step size).
This saves computational effort.
The procedure consists of ($i$) {\it lifting} (appropriate
initialization of the fine-scale simulator, an integrator of
Eq.~(\ref{microEQ1}), consistent with prescribed coarse-grained
values), ($ii$) {\it direct integration} of the microscopic
simulator during a relatively short time interval (but long enough
to accurately estimate local coarse-grained time derivatives),
($iii$) {\it restriction} (of fine-scale description onto
coarse-grained variables), and ($iv$)
{\it taking a projective step} (using a traditional numerical
integration scheme such as forward Euler).
The computational payoff of this method depends on the ratio between
a short direct integration time interval, the projective time-step
size, and the computational effort required for lifting/restriction
procedures (see e.g., Rico-Martinez et al., 2004).
More importantly, successful computation of steady states through
this method naturally attests to the {\it validity} of the chosen
coarse-grained observables in describing {\it both} fine-scale and
coarse-grained states.

\begin{figure}[t]
\begin{center}
\includegraphics[width=.8\columnwidth]{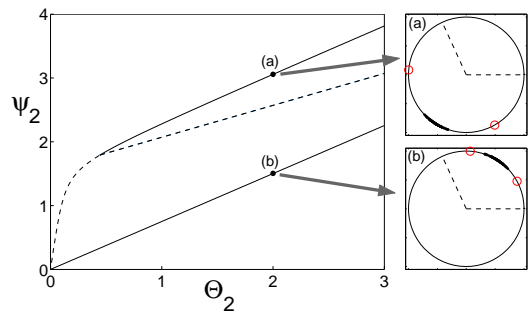}
\caption{
\label{micro}
(Left panel)
A bifurcation diagram observed on one leader $\psi_2$,
computed using AUTO2000 ($K = 1.0$).
Solid (dashed) lines represent stable (unstable) branches.
There exist a few other unstable branches that are not shown
here.
At some critical value of $\Theta_2$, an unstable state
in the upper branch undergoes a forward pitchfork bifurcation;
two unstable states coincide.
The lower branch of ``trivial'' solutions does not exhibit
any bifurcation.
(Right panels)
Snapshots of two (symmetric) stable states in the bistable
regime ($\Theta_2 = 2.0$), marked by dots in the left panel.
}
\end{center}
\end{figure}

\begin{table*}[t]
\begin{center}
\caption{\label{table}
 \label{GMRES}
A coarse steady state computation at $K = 1.0$ and
$\Theta_2 = \pi/4$ for $N = 300$, using the Newton-GMRES method.
Values at each iteration have been averaged over an ensemble of
100 realizations.
The last column shows relative residuals.
}
\begin{tabular}{ccccccc}
iteration & $\psi_1$ & $\psi_2$ & $\alpha_0$ & $\alpha_1$ & $\alpha_3$ & residuals\\ \hline
0 & 0.0 & 0.0 &0.0 & 0.0 & 0.0 & 1.0 \\
1 & 3.421$\times 10^{-5}$ & 4.143$\times 10^{-1}$ & 3.478$\times 10^{-5}$ & 5.963$\times 10^{-2}$ & 2.969$\times 10^{-9}$ & ~2.680$\times 10^{-3}$ \\
2 & 8.871$\times 10^{-4}$ & 3.900$\times 10^{-1}$ & 9.293$\times 10^{-4}$ & 8.632$\times 10^{-2}$ & 2.435$\times 10^{-5}$ & ~8.135$\times 10^{-4}$ \\
3 & 1.245$\times 10^{-3}$ & 3.969$\times 10^{-1}$ & 1.991$\times 10^{-3}$ & 9.819$\times 10^{-2}$ & 9.387$\times 10^{-5}$ & ~2.056$\times 10^{-4}$ \\
4 & 1.338$\times 10^{-1}$ & 5.275$\times 10^{-1}$ & 2.679$\times 10^{-1}$ & 1.010$\times 10^{-1}$ & 8.338$\times 10^{-4}$ & ~3.820$\times 10^{-5}$ \\
5 & 1.959$\times 10^{-1}$ & 5.896$\times 10^{-1}$ & 3.929$\times 10^{-1}$ & 1.010$\times 10^{-1}$ & 1.754$\times 10^{-4}$ & ~3.660$\times 10^{-7}$ \\
6 & 1.958$\times 10^{-1}$ & 5.896$\times 10^{-1}$ & 3.927$\times 10^{-1}$ & 1.010$\times 10^{-1}$ & 1.760$\times 10^{-4}$ & ~~6.513$\times 10^{-12}$ \\
\end{tabular}
\end{center}
\end{table*}

Projective integration using five coarse-grained
variables ($\psi_1, \psi_2$, and the first three
non-vanishing $\alpha_n$'s; $\alpha_0$, $\alpha_1$, and
$\alpha_3$) follows virtually the same trajectories of the
full, direct integration (Fig.~\ref{PI}), even if $\omega$ is
{\it newly} drawn at each lifting; the agreement is even
better if the same $\omega$ were used (hence the dynamics
are fully deterministic).
Both {\it lifting} (simply using Eq.~(\ref{expansion})) and
{\it restriction} (least squares fitting) operations require
minimal computational efforts.
Therefore, the computational efficiency in the present case is
nearly exclusively determined by the projective step size,
which is a factor of about four in Fig.~\ref{PI};
with a more sophisticated projection algorithm, a higher
efficiency can be obtained.
We see that both $\alpha_1$ and $\alpha_3$ reach their steady
state values quickly ($t \sim 5$), showing that the correlation
between $\theta$ and $\omega$ are fully developed by then.
However, the other chaos coefficient $\alpha_0$ (representing
the average direction) slowly drifts towards the steady state,
and so do $\psi_1$ and $\psi_2$ (note that it is still varying
at $t = 500$); the computation of an asymptotic, steady state
requires a very long time integration.

Direct integration (including projective integration) cannot
compute unstable steady states and are inappropriate for
stability computations and parametric bifurcation studies.
Both stable and {\it unstable} steady state values can be
systematically (and much more efficiently than the projective
integrations) computed by applying coarse-grained fixed point
algorithms to the steady state condition of
Eq. (\ref{coarse_stepper}), i.e.,
$\mbox{\boldmath $x$} - {\mathbf \Phi}_{\Delta T}(\mbox{\boldmath $x$}) = 0$,
in {\it much lower} dimension than that of individual level.
We use the coarse Newton-GMRES (Kelley, 1995), a matrix-free,
method to compute coarse-grained fixed points.
We observe that the algorithm accurately converges within
a few steps (Tab.~\ref{GMRES});
the converged values are accurately consistent with the restricted
values of the fixed point solution of the detailed
(i.e., ($N$+2)-dimensional)
problem, within prescribed convergence tolerance.
By combining a coarse fixed point algorithm with pseudo-arclength
continuation (Keller, 1987), we numerically compute coarse-grained
bifurcation diagrams in the following sections.
The computational efficiency of the coarse fixed point algorithm
varies with the choice of the initial guess for the iteration.
With a totally uneducated guess, it could take even longer than
the direct integration (note that the latter never computes the
exact solutions); however, during the continuation computation
shown below, a good initial guess is always available from the
previous parameter value(s), and even several orders of magnitude
of computational efficiency can be achieved.

\begin{figure}[t]
\begin{center}
\includegraphics[width=.6\columnwidth]{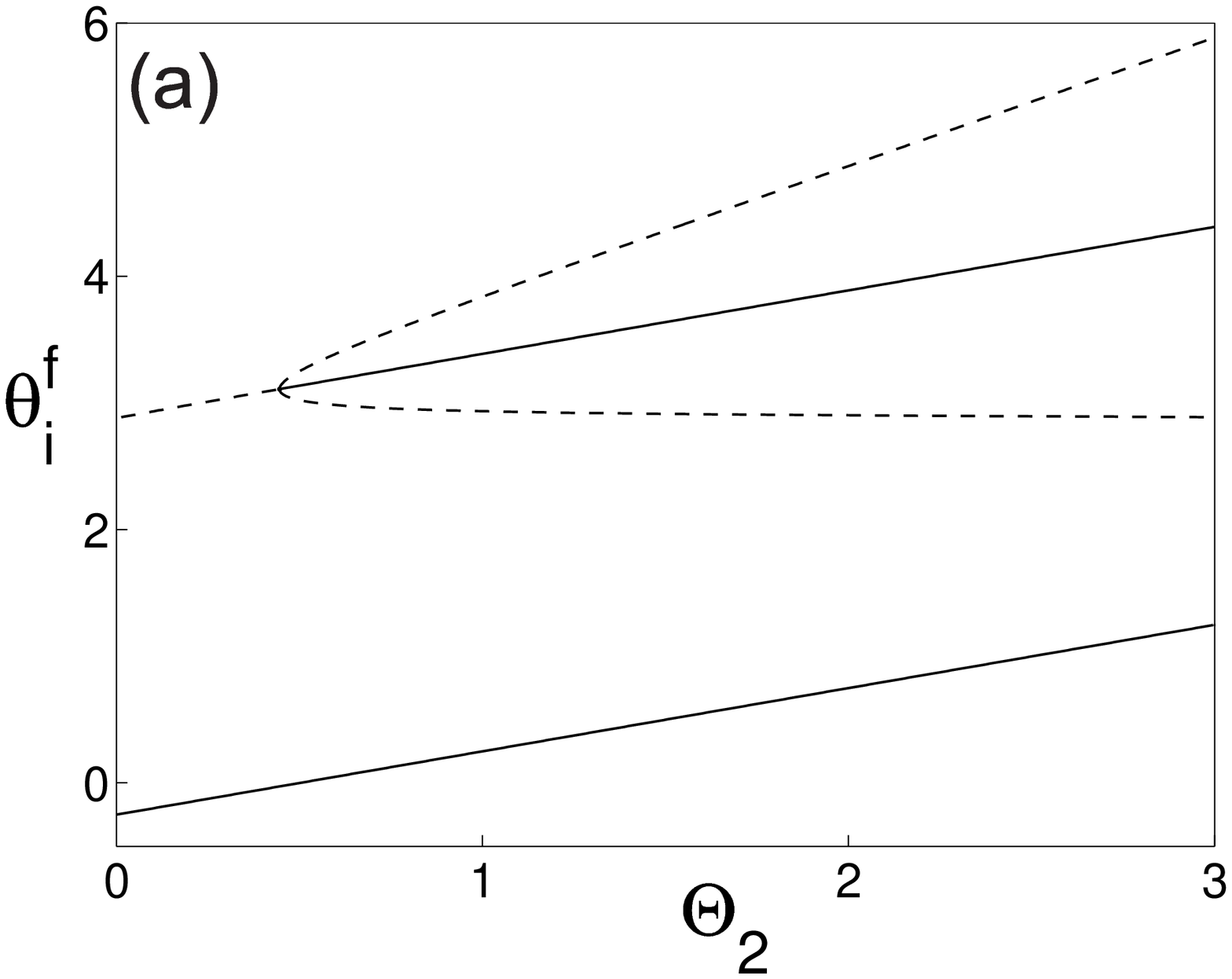}
\includegraphics[width=.6\columnwidth]{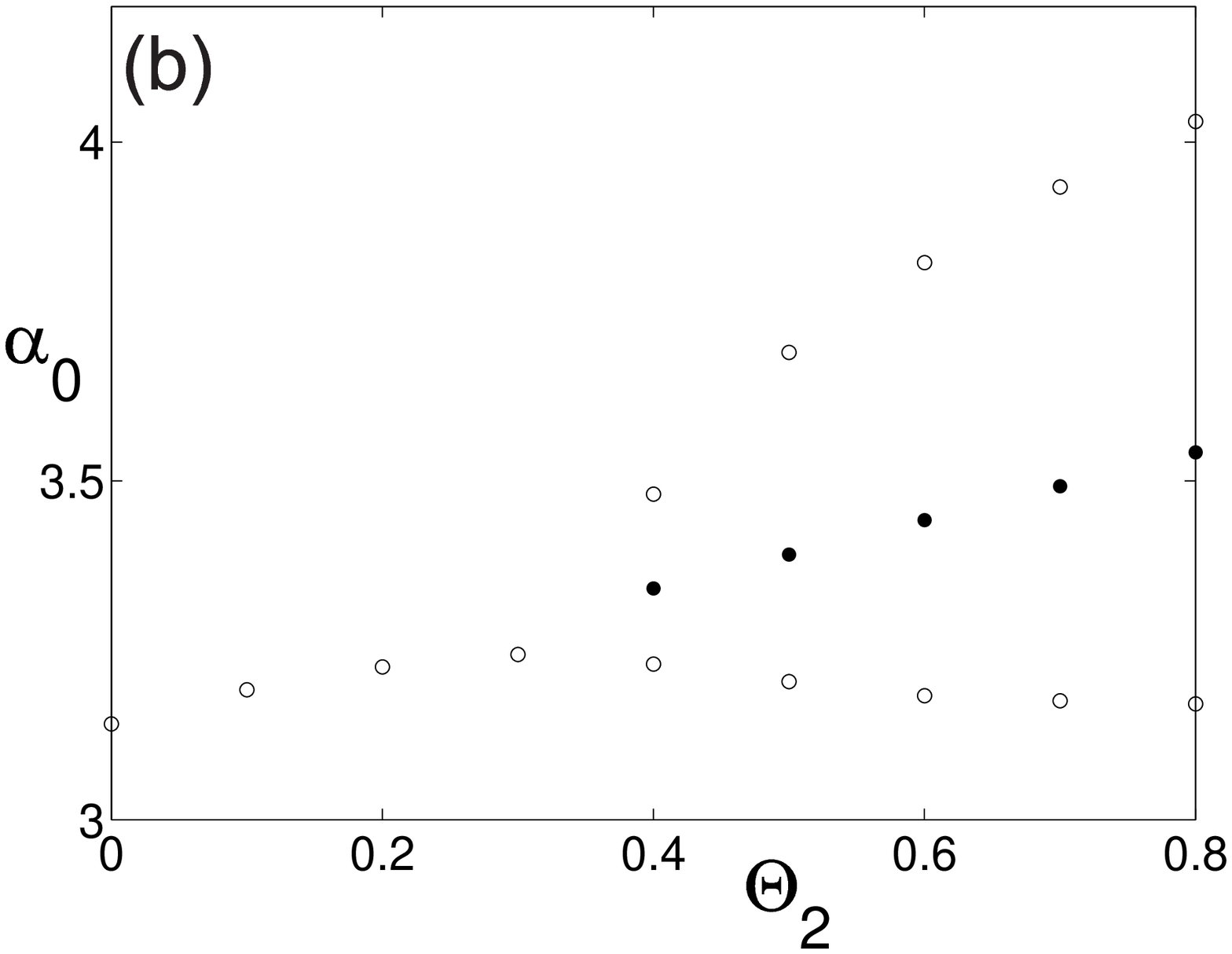}
\caption{
\label{macroTh2}
(a)
A bifurcation diagram observed on one (arbitrarily chosen)
follower, as a function of $\Theta_2$ ($K = 1.0$), computed using
AUTO2000 (the same case as in Fig.~\ref{micro}).
Superscript `f' has been added to emphasize that this is the
orientation of {\it a follower}.
A few other existing unstable branches are not included here.
The upper branch undergoes a pitchfork bifurcation and becomes
stable.
(b)
A coarse bifurcation diagram observed on $\alpha_0$ (average
direction), obtained by the coarse Newton-GMRES method with
pseudo arc-length continuation.
Only a blowup around the bifurcation point is shown.
Coarse-grained dynamics exhibit the same structure as in
the fine-scale level.
Filled (open) circles represent stable (unstable) steady
states.
}
\end{center}
\end{figure}

\subsection{Types of fine-scale dynamical behavior}

We first analyze the detailed $(N+2)$-dimensional fine-scale
model in the full synchronization regime, in order to obtain
insights on fine-scale dynamics to be compared with our
coarse-grained analysis below.
We use AUTO2000 (Doedel et al., 2000) to compute the fine-scale
bifurcation diagrams as functions of $\Theta_2$ at a fixed value
of $K$;
only projections for one leader ($\psi_2$) are shown in
Fig.~\ref{micro} and for one follower in Fig.~\ref{macroTh2} (a).
All the other followers exhibit essentially the same dynamical
behavior as the one shown here (except for some quantitative
differences).

The interaction between the individuals causes the steady
state directions of the leaders to deviate from the preferred
angles $0$ and $\Theta_2$, respectively.
Such deviation can occur in two directions, either toward the
region bounded by $[0,\Theta_2]$ (an ``obvious'' steady state
where followers are directed in between the directions of the
leaders; see Fig.~\ref{micro} (b)) or the other way around
(e.g., Fig.~\ref{micro} (a)).
The analysis shows that for small $\Theta_2$ values only the
former state is stable, while for large values, both of these
states become stable.
The branches for ``obvious'' stable steady states, which
correspond to lower straight solid lines, exhibit no
bifurcations (see Figs.~\ref{micro} and \ref{macroTh2} (a)).
On the other branches, forward pitchfork bifurcations 
at some critical value of $\Theta_2$ give birth to another
stable branch (a state on this stable branch is shown in
Fig.~\ref{micro} (a)), as well as two unstable asymmetric
solution branches, hence the population becomes bistable.
The critical value of $\Theta_2$ for the onset of the bistability
depends on $K$ (precisely speaking, $K/\sigma_{\omega}$);
the critical value is $\Theta_2 \sim 0.45~(2.2)$ at
$K = 1.0~(0.5)$.
As $K$ decreases further, the critical value monotonically
increases until fully synchronized steady states lose
stability at some critical value of $K$.

\subsection{Coarse-grained dynamics}

We now compute coarse-grained steady state solutions.
A coarse-grained bifurcation diagram for $\alpha_0$ (representing
the average direction of the followers) is compared with the
corresponding diagram observed for {\it one} follower, in
Figs.~\ref{macroTh2} (b) and (a); (b) is a blowup of the
region around the bifurcation.
Both sides of the bifurcation point can be described by
the {\it same} set of coarse-grained observables, which
clearly summarize group level dynamical behavior of the
followers before and after the bifurcation.

As $K$ decreases in the Kuramoto model, oscillators get
desynchronized (Kuramoto, 1984), starting with the oscillator
with the maximum value of $|\omega_i|$ (the ``extreme''
oscillator) through a saddle-node (actually a ``sniper'')
bifurcation on a limit cycle (Moon et al., 2006).
We expect the same type of bifurcation to occur in this model.
However, when we try to compute the coarse-grained steady states
as functions of $K$ using the previously mentioned five coarse
variables (via coarse Newton-GMRES method and pseudo arc-length
continuation, neither a bifurcation nor an unstable branch
is appropriately identified.
The computation, initialized at large $K$ steady states,
accurately follows stable branches down to some critical value
of $K$ (where the transition occurs), and then fails to converge.
Our coarse-grained observables are not sufficient to describe
the states on the ``other side'' of the bifurcation point,
as we will explain below.

\begin{figure}[t]
\begin{center}
\includegraphics[width=.6\columnwidth]{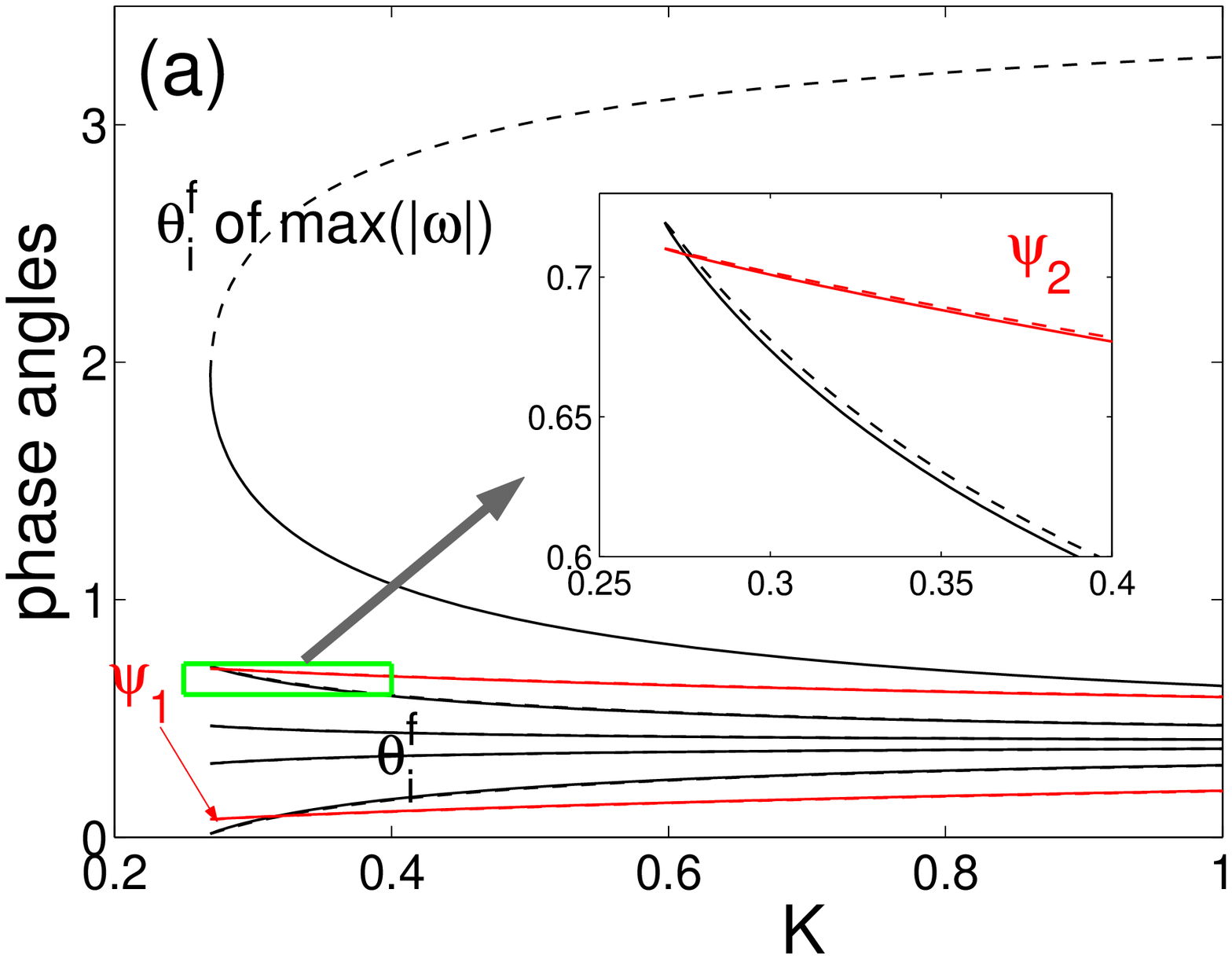}
\includegraphics[width=.6\columnwidth]{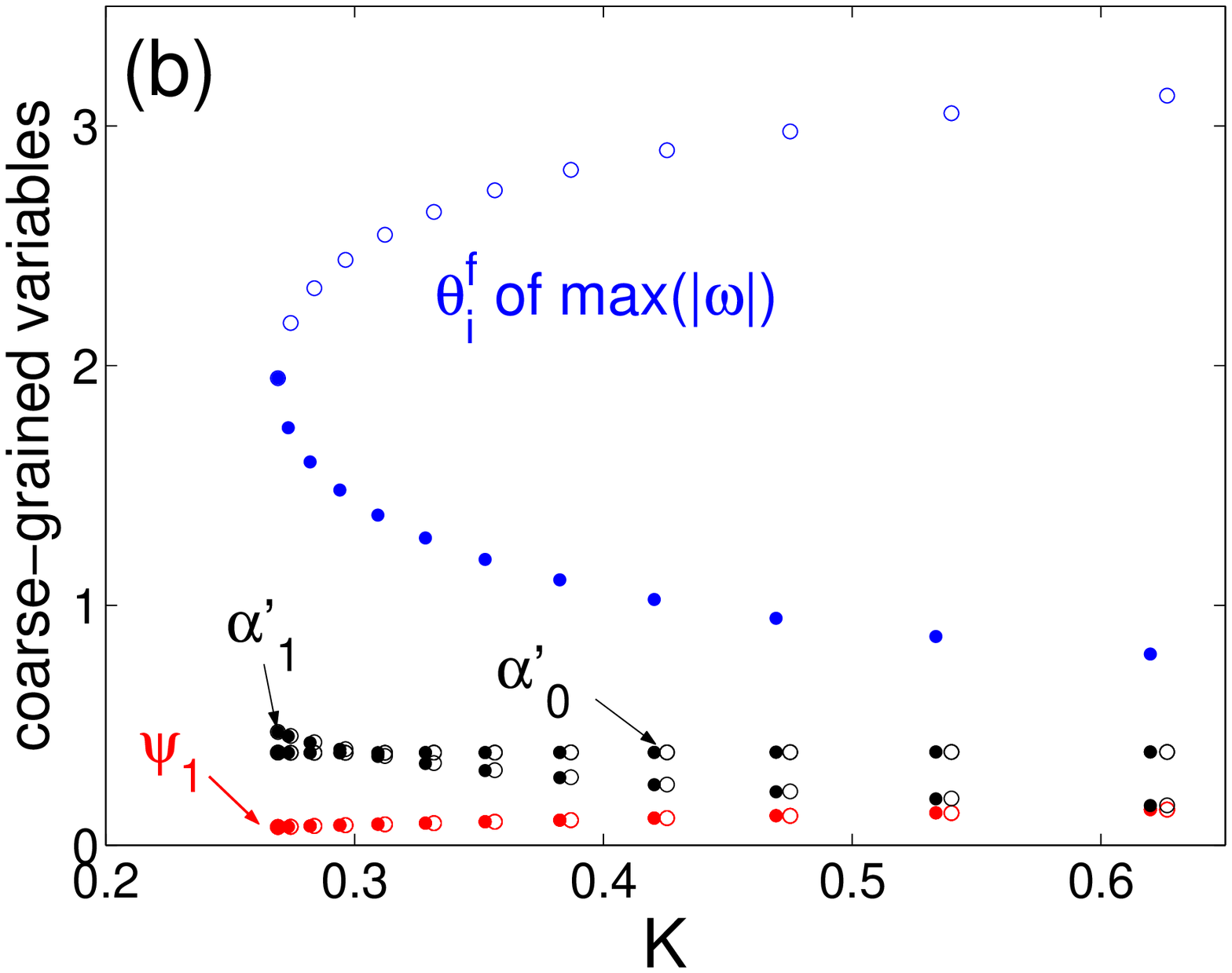}
\caption{
\label{turningPT}
(Color online)
(a)
A bifurcation diagram observed on a few followers, including the
one with the maximum of $|\omega_i|$ (the ``extreme'' follower),
as a function of $K$ ($\Theta_2 = \pi/4$), computed using
AUTO2000. 
A critical value where the extreme individual gets desynchronized
corresponds to a saddle-node bifurcation point on a limit cycle
(a ``sniper'' bifurcation).
Except for the extreme follower, stable and unstable branches
nearly coincide (see the inset).
(b)
In order to capture the fine-scale bifurcation, the angle of
the extreme follower has to be discounted from the chaos
expansion and considered as an {\it extra} coarse-grained
variable (see text).
We distinguish these chaos coefficients (from the ones used
so far) by adding a prime.
It was computed via the coarse Newton-GMRES method with
continuation.
}
\end{center}
\end{figure}

A fine-scale bifurcation diagram (computed using AUTO2000)
obtained by starting from a stable steady state on the lower
branch in Fig.~\ref{micro} is shown in Fig.~\ref{turningPT} (a).
Here the diagrams for two leaders and only a few followers,
including the extreme one, are shown.
We find that both stable and unstable branches for each
angle nearly coincide for all the individuals (see inset of
Fig.~\ref{turningPT} (a)), except for the extreme one.
As the difference between stable and unstable branches
(at the same value of $K$) is appreciable {\it only} when
observed on this extreme oscillator, a smooth mapping between
$\theta$ and $\omega$ does not prevail for unstable states,
and the previously used chaos coefficients are not appropriate
any more.

Taking these observations into account, it is easy to
remedy the situation as follows:
The fact that stable and unstable branches nearly coincide,
discounting the extreme follower, suggests that all the
individuals {\it except for the extreme follower} can be again
described by the same set of chaos coefficients.
Thus we treat the orientation of the extreme one separately
(introducing it as an additional coarse-grained variable),
and discount it from the polynomial chaos expansion.
(From the fact that the extreme follower gets desynchronized
at the transition, one can also intuitively see that followers
have to be considered as a combination of a clump of
synchronized ``bulk'' and a separate, extreme one.)
We compute the solutions with continuation, using this new
set of {\it six} coarse variables, which captures the bifurcation
and appropriately describes the unstable steady states
(Fig.~\ref{turningPT} (b)); we have analyzed exactly
the same realization used in Fig.~\ref{turningPT} (a)
for direct comparison.
When bifurcation diagrams are computed for ensembles of
many realizations, an uncertainty will arise in the exact
quantification of the bifurcation point, due to the fluctuation
of finite-dimensional random variables among realizations,
while the results are qualitatively the same as those of
a single realization (Xiu et al., 2005).

The coarse bifurcation results shown in Fig.~\ref{turningPT} (b)
illustrate that the steady state directions of the leaders and
the average direction of followers ($\alpha_0'$, discounting
the extreme one; a prime is added to distinguish it from the
previously used notation) are virtually the same for a range of $K$.
Only higher order chaos coefficients (only $\alpha_1'$ is
shown in Fig.~\ref{turningPT}) appreciably vary as a function
of $K$, which means that individuals spread more widely as $K$
decreases, until the extreme one eventually starts to oscillate
freely, while the average steady state direction remains the same.

\begin{figure}[t]
\begin{center}
\includegraphics[width=.6\columnwidth]{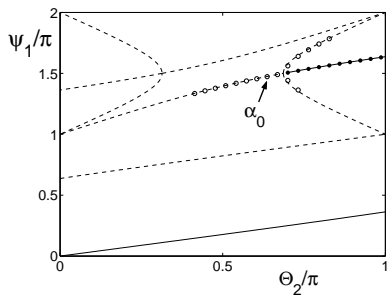}
\caption{
\label{microK2.4}
Lines: A bifurcation diagram observed on $\psi_1$ in the minimal model
(the first two ODEs in Eq.~(\ref{reduced}) with $N_1 = N_2 = 1,
~N_3 = 0$),
for varying $\Theta_2$ at a fixed value of $K = 2.4$, obtained
by AUTO2000.
For large enough preferred angles ($\Theta_2/\pi > \sim 0.7$),
the system becomes bistable through a forward pitchfork bifurcation.
Circles: A coarse bifurcation diagram near the pitchfork bifurcation,
observed on $\alpha_0$, as a function of $<\Phi>/\pi$,
computed by the coarse Newton-GMRES method.
Filled (open) circles represent stable (unstable) steady states.
}
\end{center}
\end{figure}

\section{Results for Case II}
 \label{results2}

\subsection{Dynamics of statistically similar groups}
 \label{results2a}

Here we explore both the fine-scale and coarse-grained dynamics
of a model for two groups of {\it heterogeneous} leaders (with no
followers) shown in Eq.~(\ref{for_many}), and compare the results
of the two different scales.
One notable difference from the Kuramoto model is that
``oscillators'' in Eq.~(\ref{for_many}) do not have finite
random variables (natural frequencies), hence there is no onset of
the synchronization
that occurs at a finite value of $K$ (or, they can be alternatively
seen as Kuramoto-like oscillators of zero natural frequencies,
which result in the onset at $K = 0$, hence they get synchronized
for all $K$ values).
The analysis of the minimal model (the first two of
Eq.~(\ref{reduced}) with $N_1 = N_2,~N_3 = 0$) reveals that for large
enough $\Theta_2~(>\sim \pi/2)$ the system exhibits bistability
for a certain range of $K$ (Nabet et al., 2006), as in the previous
case in Sec.~\ref{results}.
Here we will vary $<\Phi>$ as the main parameter for two
different values of $K$.
For large coupling strengths ($K > 2.0$), the bistability in the
minimal model appears through a forward pitchfork bifurcation,
when $\Theta_2$ is varied as a parameter (Fig.~\ref{microK2.4}).
This minimal model can be seen as a special case of the current
model, where both ${\mathcal X}$ and $\Phi$ are assumed to be
delta functions and each group consists of identical individuals.

\begin{figure}[t]
\begin{center}
\includegraphics[width=.6\columnwidth]{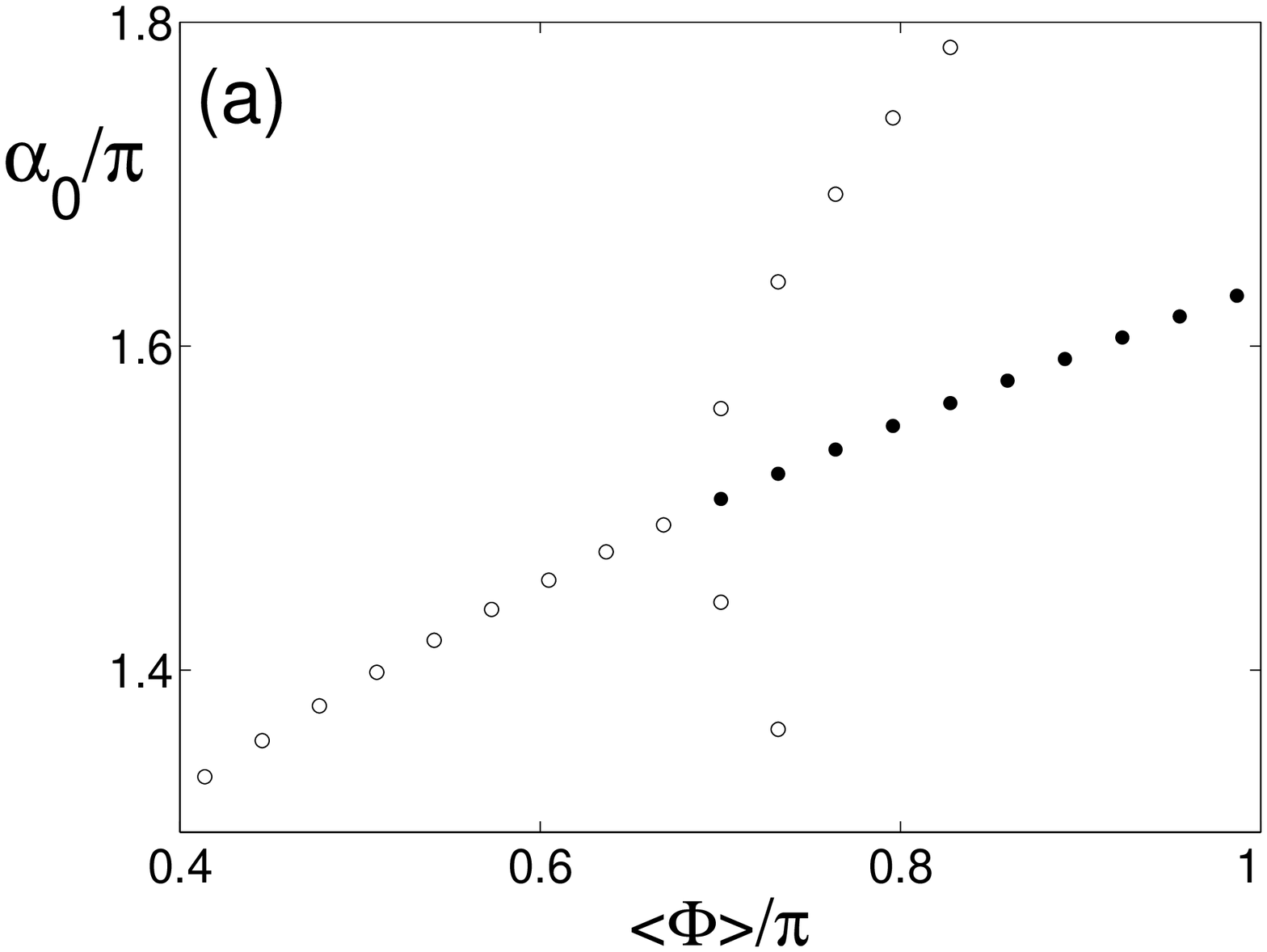}
\includegraphics[width=.6\columnwidth]{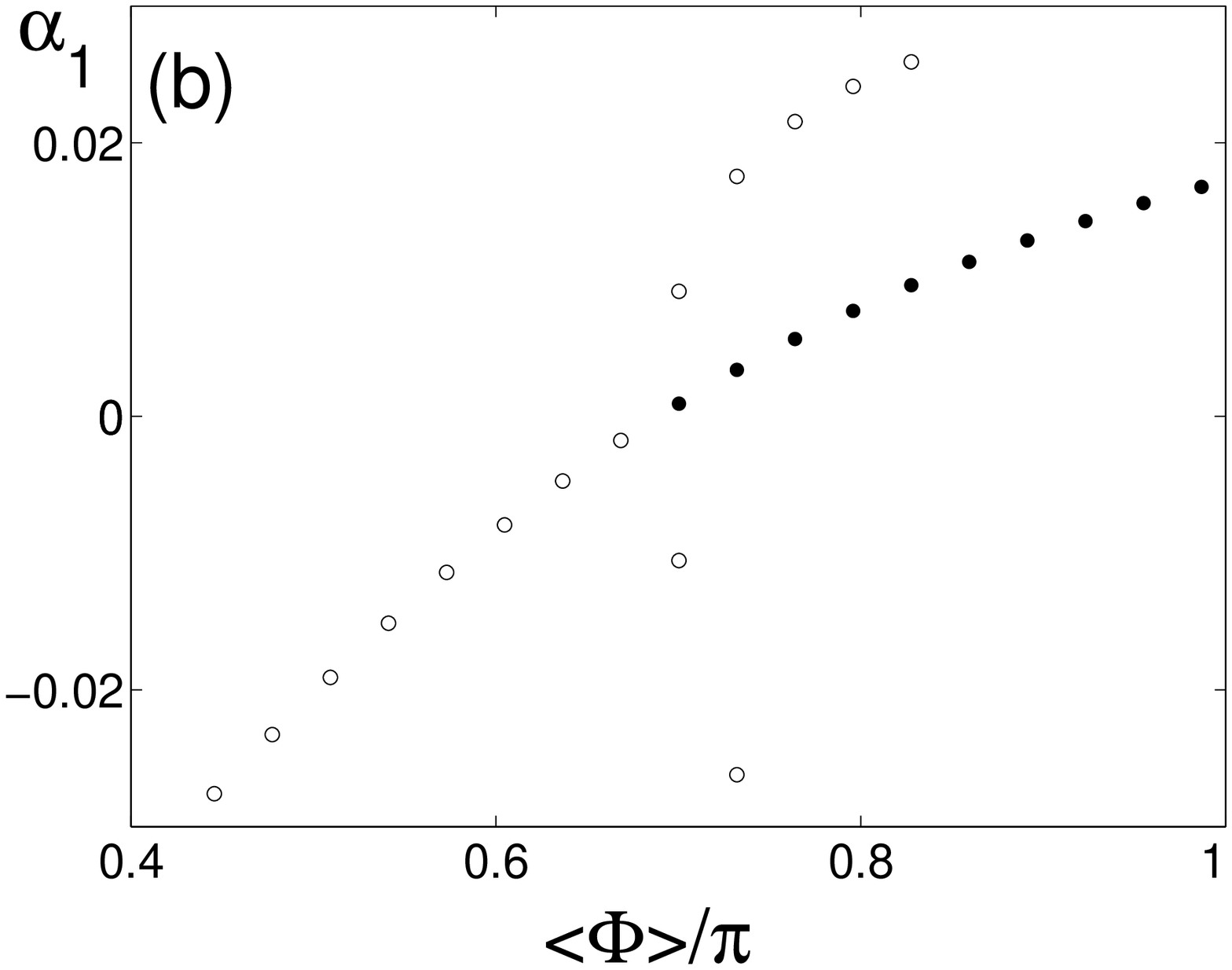}
\caption{
\label{macroK2.4}
Coarse-grained bifurcation diagrams observed on the first two chaos
coefficients: (a) $\alpha_0$; the average direction of the first group
of leaders, and (b) $\alpha_1$; the ``slope'' between $\chi$ and
${\mathcal X}$, as functions of $<\Phi>$.
These are blowups of the region around the forward pitchfork
bifurcation point in Fig.~\ref{microK2.4}.
}
\end{center}
\end{figure}

\begin{figure}[t]
\begin{center}
\includegraphics[width=.6\columnwidth]{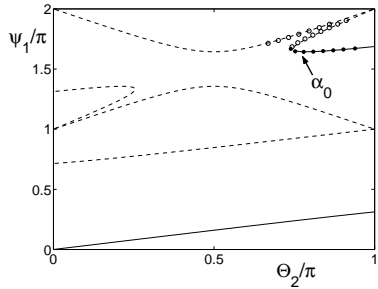}
\caption{
\label{micromacroK1.8}
Lines:
A bifurcation diagram for the minimal model of two leaders, for
varying $\Theta_2$ at a fixed value of $K = 1.8$, obtained by
AUTO2000.
For large enough preferred angles ($\Theta_2/\pi > \sim 0.75$),
the system becomes bistable, but the nature of the bifurcation is
different from that of higher $K$ value cases (a saddle-node
vs. a pitchfork bifurcation; see Fig.~\ref{microK2.4}).
Circles:
A coarse bifurcation diagram observed on the average direction of
the first group of leaders ($\alpha_0$) around the saddle-node
bifurcation, as a function of $<\Phi>/\pi$,
computed via the coarse Newton-GMRES method with continuation.
Filled (open) circles represent stable (unstable) steady states.
}
\end{center}
\end{figure}

We begin by asking whether our model for heterogeneous groups
exhibits similar types of dynamical behavior.
One can also do accelerated computations of steady states
using the coarse projective integration, but here we skip such
computations and present only the coarse bifurcation analysis
results.
{\it Coarse} bifurcation diagrams obtained through the coarse
Newton-GMRES method (Kelley, 1995) and pseudo arc-length
continuation (Keller, 1987) (for Gaussian distributions of
${\mathcal X}$ and $\Phi$; $\sigma_{{\mathcal X}} = \sigma_{\Phi} = 0.1$,
$N_1 = N_2 = 100$) show that the heterogeneous groups indeed
exhibit the same qualitative type of coarse dynamical behavior
around the pitchfork bifurcation point (Fig.~\ref{macroK2.4}).
As we consider symmetric unimodal distribution functions, all
the even order chaos coefficients (except for $\alpha_0$ and
$\beta_0$) virtually vanish.
The diagram for $\alpha_0$ of the first group $<\chi>$
(average direction) exhibits reasonably good quantitative
agreement with the corresponding diagram for the minimal model,
within fluctuations of finite-size random variables, shown
in Figs.~\ref{microK2.4} and \ref{micromacroK1.8}.
It is interesting to note that at the critical point, all
the followers are headed for the same direction ($\alpha_1 = 0$,
which corresponds to the ``slope'' between ${\mathcal X}$ and $\chi$);
see Fig.~\ref{macroK2.4} (b).

\begin{figure}[t]
\begin{center}
\includegraphics[width=.62\columnwidth]{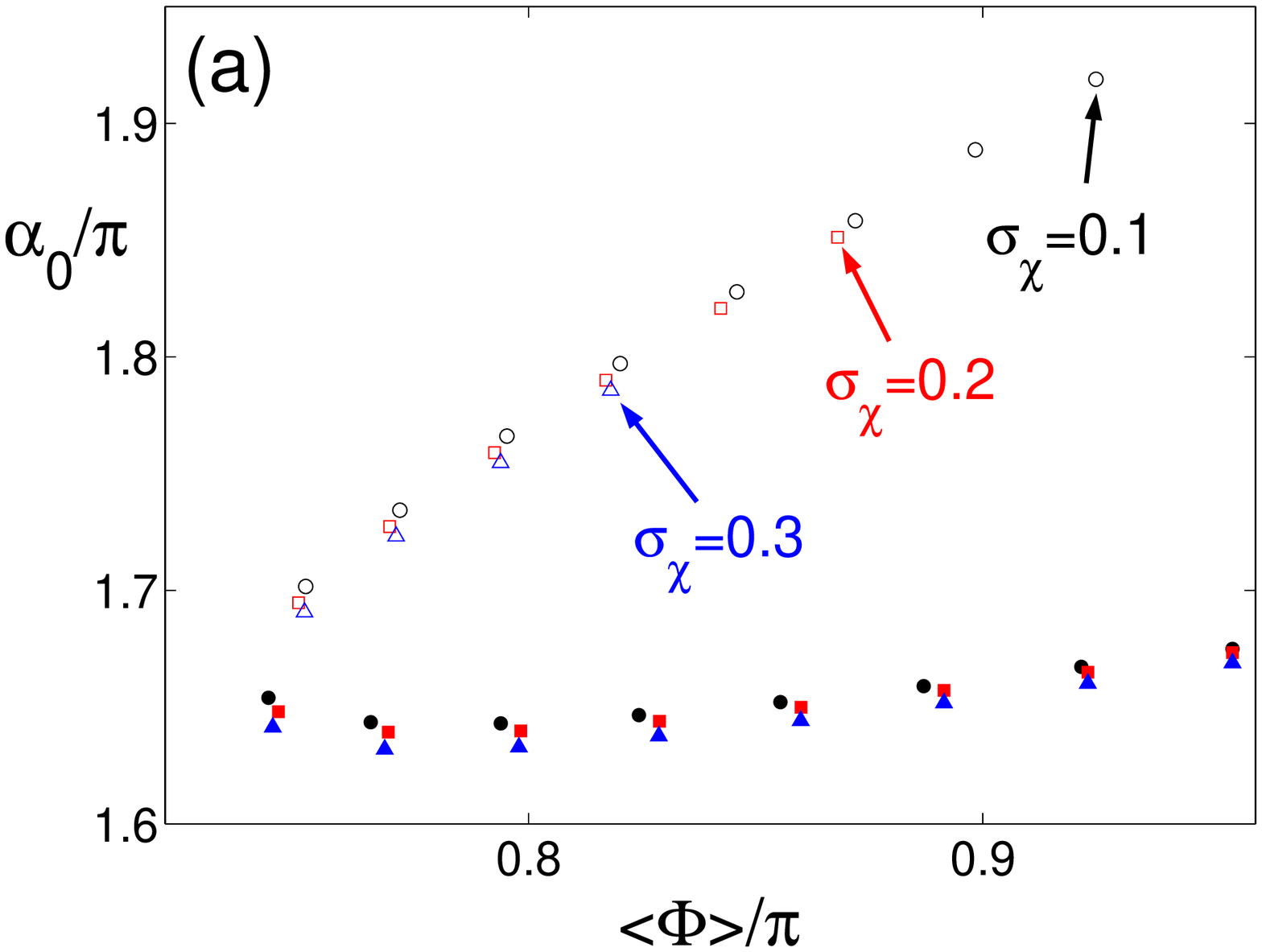}
\includegraphics[width=.6\columnwidth]{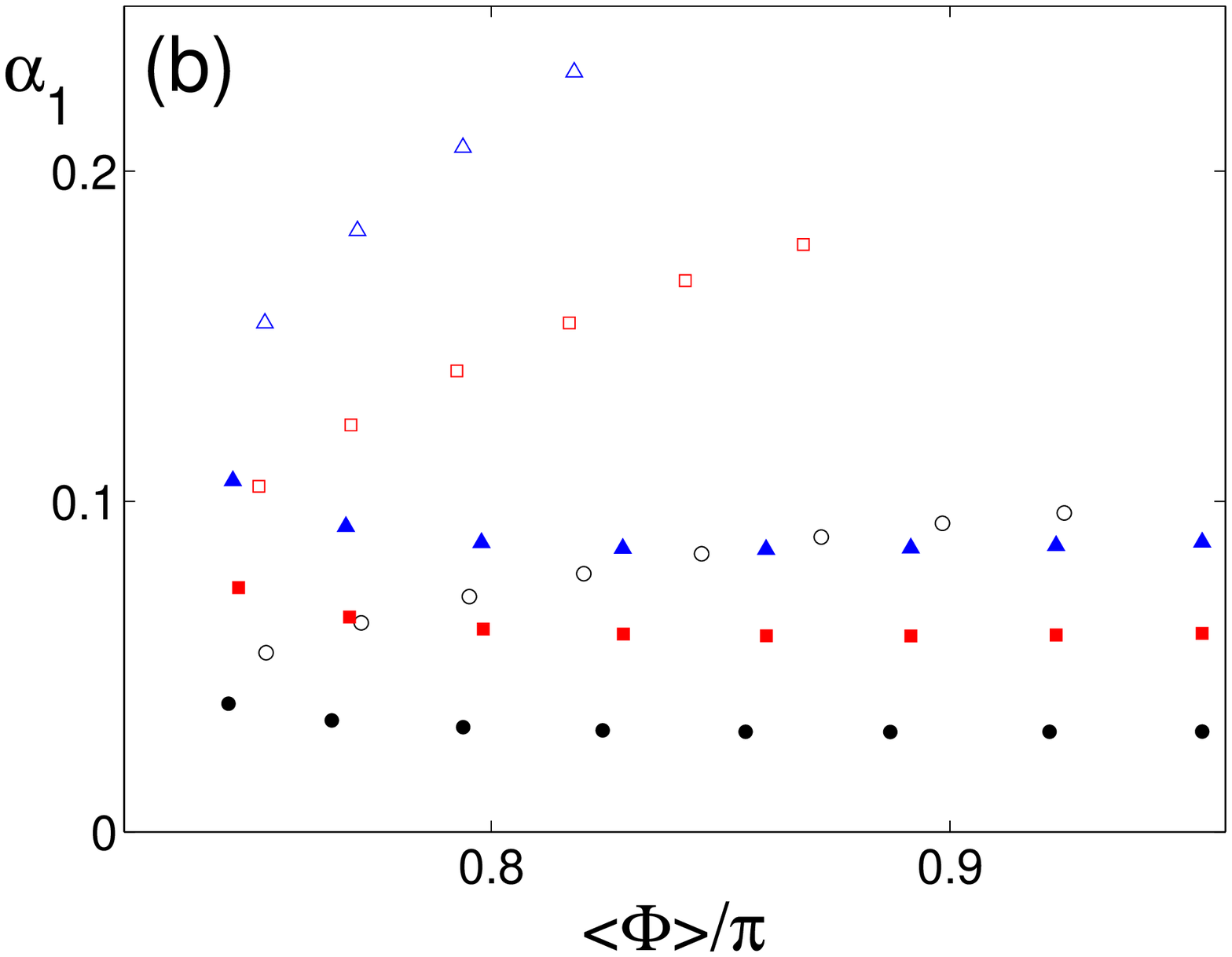}
\caption{
\label{vary_std}
(Color online)
Coarse-grained bifurcation diagrams near a turning point in
Fig.~\ref{micromacroK1.8}, for ${\mathcal X}$ distributions of three
different widths (the standard deviation $\sigma_{{\mathcal X}}$ = 0.1
for circles; 0.2 for squares; 0.3 for triangles), obtained
via the coarse Newton-GMRES method and continuation.
The standard deviation for the second group, $\sigma_{\Phi}$,
is kept the same at 0.1 ($K = 1.8,~N_1 = N_2 = 100$).
Filled and open symbols represent stable and unstable states,
respectively.
(a)
The first chaos coefficients $\alpha_0$ (average direction of
the first group) are nearly the same for the three cases. The
difference between the cases becomes apparent in higher order
coefficients that reflect the degree of spreading;
see $\alpha_1$ in (b).
}
\end{center}
\end{figure}

The Hermite polynomial expansion converges so quickly that the
expansions can be accurate even when truncated at the third order.
Due to the reflection symmetry (about $<\Phi>$/2),
\mbox{\boldmath $\beta$} coefficients have similar structures as
the \mbox{\boldmath $\alpha$} ones, after proper reflection and
translation.
Only results on \mbox{\boldmath $\alpha$} are presented here.
As the coupling strength decreases across $K = 2.0$, the
nature of the bifurcation changes (from a pitchfork) to
a saddle-node bifurcation (Fig.~\ref{micromacroK1.8})
at $K = 2.0$,
which also occurs in the model for homogeneous populations;
the nature of the transition between these different bifurcations,
a higher codimension bifurcation, has been discussed in
Nabet et al. (2006).

\subsection{Statistically different groups}
 \label{results2b}

So far we have considered statistically similar groups,
namely $N_1 = N_2$ and $\sigma_{{\mathcal X}} = \sigma_{\Phi}$;
they differed only by average preferred directions.
It is natural to ask how the dynamics change as the parameters
concerned with the distributions (for the preferred directions)
are varied.
It is readily expected that the essential dynamics of two
different-size groups can be reflected in the minimal model
using two different coupling strengths, which is considered
in Nabet et al. (2006).
Here we consider only the cases with varying width of the
distributions ($\sigma_{\Phi} \neq \sigma_{{\mathcal X}}$), which has
no analog in the minimal model.

Coarse bifurcation diagrams for three different Gaussian
distributions for ${\mathcal X}$ ($\sigma_{{\mathcal X}}$ is varied while
$\sigma_{\Phi}$ is kept at 0.1; see Fig.~\ref{vary_std}) show
that the average directions ($\alpha_0$'s) hardly vary with
the width of the distributions; the primary parameter that
affects on the average direction is the {\it group size}.
For the distributions of different widths, the fixed point
computation with continuation fails to converge at different
values of $\alpha_0$'s;
points marked by arrows in Fig.~\ref{vary_std} are the last points
the Newton-GMRES computations converged in each case, when
approached from the stable branches.
Such a failure of convergence can be expected, because the steady
states on this unstable branch overlap with another nearby unstable
branch (which is not shown in this figure, but was shown in
Fig.~\ref{micromacroK1.8});
characterizing the distribution with a few Wiener chaos
coefficients does not provide an accurate description any more.
The differences between the three cases (of different distribution
widths) manifest themselves clearly in higher order chaos coefficients.
While the average behavior remains nearly the same (Fig.~\ref{vary_std}
(a)), individuals in the group spread more widely (as reflected in
$\alpha_1$ and higher order coefficients; Fig.~\ref{vary_std} (b)),
as the width of the random variable (distribution) increases.

\section{Conclusions}
 \label{conclusions}

We have demonstrated a computational venue (an equation-free
polynomial chaos approach) to study coarse-grained dynamics
of individual-based models accounting for the {\it heterogeneity}
among the individuals in animal group alignment models.
We considered {\it finite} populations of (I) two ``leaders''
(which have direct knowledge on preferred directions) and
$N (\gg 1)$ uninformed, heterogeneous ``followers'', and
(II) two groups of heterogeneous ``leaders''.
We explored the {\it coarse-grained}, group level
(low-dimensional) dynamics using the polynomial chaos expansion
coefficients as coarse-grained observables;
these observables account for rapidly developing correlations
between random variables, and sufficiently specify both
fine-scale and coarse-grained (group-level) dynamical states.

All the analysis in our study was done expressively avoiding
the derivation of coarse-grained governing equations, following
a {\it nonintrusive}, equation-free computational approach
wrapped around the direct system simulator.
It should be noted that
we have not assumed that $N$ is infinitely large (so-called
the ``continuum limit'').
Our approach can be used for systems of any finite, large
number of populations, and it can be equally applied to various
types of random variables (following generalized polynomial
chaos) and/or various heterogeneity.
We compared our results with those of minimal models that do
not account for heterogeneity among the individuals.
They show good agreement in the lowest order (i.e., average
directions), which clearly highlights the correspondence
between the individual- and group-level dynamics
(Figs.~\ref{microK2.4} and \ref{micromacroK1.8}).
Indeed this implies that the results in Nabet et al. (2006),
where no heterogeneity is explicitly accounted for, are more
robust than demonstrated in that paper alone.

In order to analyze different coarse-grained bifurcations,
it became necessary to use different sets of coarse-grained
variables, even if the model is {\it the same}
in the fine-scale level (Fig.~\ref{turningPT}).
This clearly shows that an appropriate choice of coarse-grained
observables (in terms of which one can obtain useful closures)
is an essential step;
different coarse-grained observables are required, as the same
fine-scale model closes differently.

In the present study, we assumed that the orientational
dynamics can be separated from their translational counterpart,
and considered the simplest nontrivial cases of all-to-all
(``all-visible''), sinusoidal coupling.
Our future work will involve the incorporation of the translational
dynamics and more complicated coupling/network topology, including
heterogeneous couplings.
Our work presented here is the first step of our effort toward
the development of more detailed (and biologically more
plausible) models and their {\it coarse-graining}.

\section*{Acknowledgements}

S.J.M. and I.G.K. were financially supported by DOE and
NSF grant EF--0434319.
S.A.L. was supported in part by NSF grant EF--0434319 and
DARPA grant HR0011--05--1--0057.
B.N. and N.E.L. were supported in part by ONR grants
N00014--02--1--0826 and N00014--04--1--0534.

\newpage

\end{document}